\documentclass[]{interact}
\usepackage{lipsum}  
\usepackage{epstopdf}% To incorporate .eps illustrations using PDFLaTeX, etc.
\usepackage{xcolor}
\usepackage[natbibapa,nodoi]{apacite}% Citation support using apacite.sty. Commands using natbib.sty MUST be deactivated first!
\usepackage{graphicx}% Include figure files
\usepackage{float}
\graphicspath{{PNG/} }
\usepackage{comment}
\usepackage{algorithm}
\usepackage{algorithmic}
\usepackage{array}
\usepackage{multirow}

\theoremstyle{plain}% Theorem-like structures provided by amsthm.sty

\theoremstyle{definition}

\theoremstyle{remark}

\usepackage{graphics}
\usepackage{color}
\definecolor{rev1}{rgb}{0,0,0}

\begin{document}

%\articletype{ARTICLE TEMPLATE}% Specify the article type or omit as appropriate

%\title{Nonlinear proper orthogonal decomposition surrogate modeling of Rayleigh Bénard convection}

%This paper has been passed through a check-speller and a grammatical check.

\title{\textcolor{rev1}{Numerical assessments of a nonintrusive surrogate model based on recurrent neural networks and proper orthogonal decomposition: Rayleigh Bénard convection}}

\author{
\name{Saeed Akbari\textsuperscript{a}, Suraj Pawar\textsuperscript{a} and Omer San\textsuperscript{a}\thanks{CONTACT O. San. Email: osan@okstate.edu}}
\affil{\textsuperscript{a}School of Mechanical \& Aerospace Engineering, Oklahoma State University, Stillwater, OK 74078, USA.}
}

% \author{
% \name{Saeed Akbari\textsuperscript{a}, Omer San\textsuperscript{a}\thanks{CONTACT O. San. Email: osan@okstate.edu}, and Adil Rasheed\textsuperscript{b}}
% \affil{\textsuperscript{a}School of Mechanical \& Aerospace Engineering, Oklahoma State University, Stillwater, OK 74078, USA.}
% \affil{\textsuperscript{b}Department of Engineering Cybernetics, Norwegian University of Science and Technology, N-7465, Trondheim, Norway.}
% }

\maketitle

\begin{abstract}
Recent developments in diagnostic and computing technologies offer to leverage numerous forms of nonintrusive modeling approaches from data where machine learning can be used to build computationally cheap and accurate surrogate models. To this end, we present a nonlinear proper orthogonal decomposition (POD) framework, denoted as NLPOD, to forge a nonintrusive reduced-order model for the Boussinesq equations. In our NLPOD approach, we first employ the POD procedure to obtain a set of global modes to build a linear-fit latent space and utilize an autoencoder network to compress the projection of this latent space through a nonlinear unsupervised mapping of POD coefficients. Then, long short-term memory (LSTM) neural network architecture is utilized to discover temporal patterns in this low-rank manifold. While performing a detailed sensitivity analysis for hyperparameters of the LSTM model, the trade-off between accuracy and efficiency is systematically analyzed for solving a canonical Rayleigh-Bénard convection system.
%Within the area of computational fluid dynamics, there is always a trade-off between the available computational resources and the desired level of accuracy. In recent years, we have been steadily moving away from sparse data to rich data regimes in fluid dynamics, thanks to rapid advances in diagnostic and computing technologies. These developments offer to leverage numerous forms of nonintrusive modeling approaches from data where machine learning can be used to build computationally cheap and accurate surrogate models. To this end, we present a nonlinear proper orthogonal decomposition (POD) framework, denoted as NLPOD, to forge a nonintrusive reduced-order model approach for considering the system of Boussinesq equations. In our NLPOD approach, we first employ the POD procedure to obtain a set of global modes to build a linear-fit latent space and utilize an autoencoder network to compress the projection of this latent space through a nonlinear unsupervised mapping of POD coefficients. Then, long short-term memory (LSTM) neural network architecture is utilized to discover temporal patterns in this low-rank manifold. While performing a detailed sensitivity analysis for hyperparameters of the LSTM model, the trade-off between accuracy and efficiency is systematically analyzed for solving a canonical Rayleigh-Benard convection (RBC) system.
\end{abstract}


%%%%%%%%%%%%%%%%%%%%%%%%%%%%%%%%%%%%%%%%%%%%%%%%%%
\section{Introduction}
\label{intro}
%%%%%%%%%%%%%%%%%%%%%%%%%%%%%%%%%%%%%%%%%%%%%%%%%%
%1. Reduced order modeling 
%=============================
Within the area of computational sciences and engineering, there is always a trade-off between the available computational resources and the desired level of accuracy. In recent years, we have been steadily moving away from sparse data to rich data regimes in many disciplines, thanks to rapid advances in diagnostic and computing technologies. Consequently, various areas of science and engineering employ reduced order modeling (ROM) to overcome computational burden. For example, ROM attracts the attention of computational scientists in the fields of data assimilation, control systems, design optimization, uncertainty quantification, and sensitivity analysis as they all require a large number of simulations. In this regard, the area of fluid mechanics has access to a strong mathematical model through Navier-Stokes equations enabling researchers to build inexpensive projection-based surrogate models that can be used for multi-query tasks or performing simulations in a short amount of time to make near real-time decisions. These surrogate models are often considered as key enablers toward next-generation digital twins and computational workflows \citep{kapteyn2021probabilistic,san2021hybrid,ahmed2021closures,vinuesa2020role,rasheed2020digital,brunton2020machine}.

Broadly speaking, ROMs \textcolor{rev1}{aim at} capturing the behaviour of complex physical phenomena with a low but acceptable resolution by observing available data, describing the physical features.
%which describes the physical Phenomena, in order to predict their future behaviour.
\textcolor{rev1}{In this regard, ROMs are the engines of such multi-query workflows in converting offline data collection, processing, and training time to online execution and inference as needed.
We often seek for the temporal and spatial variations of the desired physical quantities while measuring and evaluating them, which is a crucial element to keep in mind when studying physical phenomena. 
Therefore, redundant data that barely affect dynamics cannot be regarded as descriptive.
%An important point to note in the study of physical phenomena is that we are looking for the temporal and spatial variations of desired physical quantities when measuring and analyzing them. 
%Therefore, redundant data whose effects on dynamics are negligible does not help the analysis and cannot be considered as descriptive. 
Instead, the measurement frame should be located at the times and locations where the most significant changes in the system take place, thus this data explains the nature of the physical phenomenon. 
%This is implying that not all data snapshots are worth maintaining and this is the point that should be considered in designing ROMs at the first step.
This implies that not all data snapshots are valuable enough to keep, and this is the issue that needs to be taken into account while developing ROMs.}
ROMs are categorized as intrusive and nonintrusive based on their dependence on governing equations.

Intrusive ROMs often utilize the underlying partial differential equations (PDEs) to describe the dynamics in a reduced subspace. One of the most popular \textcolor{rev1}{techniques} is the Galerkin approach, where the underlying PDEs are projected onto a set of basis functions
\citep{kalb2007intrinsic,bergmann2009enablers,esfahanian2015simulation}.
On the other hand, a nonintrusive ROM (NIROM) does not depend on governing equations, and
it gives scientists and engineers considerable flexibility when they deal with complex problems where the detailed governing equations might be out of reach, \textcolor{rev1}{or some parts of which are not fully known. As a key enabler,}
%In this regard, POD in an established ROM, which has been widely used to model fluid flows.
%2. POD
%=============================
proper orthogonal decomposition (POD), introduced by \citet{lumley1967structure}, has been one of the most common linear model reduction approaches that can be combined with the Galerkin projection to make an intrusive model. Importantly, POD can be also used together with time series prediction tools to build nonintrusive models. The underlying idea of POD has received multiple revisits by \citet{pearson1901liii,kosambi1943statistics,karhunen1946spektraltheorie,loeve1948functions,pugachev1953general,obukhov1954statistical} to become mature before introducing to the fluid dynamics community. The snapshots based POD, established by \citet{sirovich1987turbulence}, has become popular in the fluid dynamics field
\citep{deane1991low,aubry1993preserving,berkooz1993proper,park1998efficient,holmes1997low,kunisch1999control,christensen1999evaluation,ravindran2000reduced,volkwein2001optimal,kunisch2001galerkin,rathinam2003new,burkardt2006centroidal,burkardt2006pod,holmes2012turbulence}. In this method, data is stored in a matrix whose number of columns is the number of temporal snapshots. Every column is a vector representing flow field data for the whole geometry, which might be flattened in case of geometries with more than one dimension, at specific times.
%orthogonal modes.. optimal modes in the least square sense when we consider a linear subspace. 
%3. Kolmogorov barrier things. To capture underlying dynamics, we often need to include many modes.. On the hand, ML offers alternative nonlinear model reduction approaches.  Data-driven reduced order modeling (nonintrusive) ..   importance and advances in ML
%=============================
Although POD has maintained superiority in building low dimensional subspace with the least possible bases, this superiority has been maintained \textcolor{rev1}{only in linear spaces.} High nonlinearity of convection-dominated flows, such as the Rayleigh-Benard convection (RBC), causes a considerable projection error between the reconstructed data and true solution. This limitation of linear-based methodologies in representing the underlying solution manifold is often denoted as the ``Kolmogorov barrier" \citep{kolmogoroff1936uber,greif2019decay,ahmed2020breaking}. %Generally, POD is able to compress data with low number of modes for dissipative flows and with high number of modes for convection-dominated flows.
\textcolor{rev1}{Generally, POD is able to compress data with low number of modes for dissipative or time periodic flows, while it might face challenges for convection-dominated or irregular patterned flows.}
%+++ Add equations for Kolmogorov barrier and explain it in detail

To remove the projection error, either a large number of POD modes or breaking nonlinear correlations are required. Localization-based techniques have been employed to cross the Kolmogrov barrier by building multiple local subspaces in the parameter space \citep{eftang2010hp,eftang2011hp,haasdonk2011training,eftang2012parameter}, in time \citep{dihlmann2011model,drohmann2011adaptive,san2015principal,ahmed2018stabilized}, in solution features \citep{redeker2015pod}, or in state-spaces
\citep{amsallem2012nonlinear,washabaugh2012nonlinear,peherstorfer2014localized,amsallem2015fast,wieland2015implicit,grimberg2021mesh} to reduce the projection error.

%=============================
On the other hand, machine learning (ML) offers alternative nonlinear model reduction approaches. For instance, \citet{kaiser2014cluster,amsallem2016pebl,shahbazi2019reduced} have utilized k-mean algorithm, which is an unsupervised learning technique in classical ML, to cluster the snapshots and construct local subspaces. With the abundance of data acquired from numerical simulations and experiments, powerful computational resources, and user friendly libraries (e.g., PyTorch, TensorFlow), fluid dynamicists have found considerable interest in deep neural networks (DNNs) to tackle computational bottlenecks. Along these lines, \citet{pawar2019deep} have built a surrogate model for complex fluid flow utilizing POD to compress data and DNN to forecast dynamics of the system. On the one hand, they have found that the POD-DNN technique delivers accurate and stable predictions while the solution of the Galerkin projection is unstable given small number of POD modes for a highly nonlinear convection-dominated fluid flow. On the other hand, the Kolmogorov barrier has not been lifted by their investigation. \citet{srinivasan2019predictions} \textcolor{rev1}{have} compared DNN and long short-term memory (LSTM) architectures for modeling turbulent shear flows. They have found that even though both are able to capture flow structures, LSTM prediction gives lower error for forecasting both turbulence statistics and dynamics.

In order to overcome POD limitations on advection dominated flows, \citet{wang2016deep} have developed an autoencoder (AE) network for dimensionality reduction. Their method has had capability of reconstructing solution space with lower mean squared error than POD given the same number of modes. In other words, the nonlinear autoencoder technology is able to transform data to a lower dimensional space than POD with same reconstruction error.
\citet{eivazi2020deep} have combined a nonlinear multi-layer perceptron (MLP) based autoencoder model with the power of an LSTM neural network to compress high fidelity data to a latent space and forecast future states of low fidelity data. Since the MLP networks use one neuron for each computational node and all the neurons are fully connected in this network, the number of weights can quickly explode, and consequently, a large amount of memory is required to manage this network. Moreover, MLP architectures are not translation invariant and they are not able to extract features in a non-separable space. On the other hand, \citet{maulik2021reduced} have incorporated convolutional autoencoder (CAE) with LSTM and have employed the capability of preserving translation invariance and extracting features in a non-separable space of convolutional neural networks. Convolution operator causes a flow field to lose information in corners, so padding is employed to preserve information in the corners by adding extra nodes. Since the values of desired quantity are not known beyond the boundaries, CAE might face challenges in reconstructing correct values at the boundaries. 
%Besides, a common CAE cannot be applied to a flow field with unstructured mesh. 
In their recent work, \citet{ahmed2021nonlinear} have developed nonlinear proper orthogonal decomposition \textcolor{rev1}{(NLPOD)} by combining POD and AE \textcolor{rev1}{to reduce the number of degrees of freedom needed to represent the underlying dynamics.} They have successfully used only two modes in latent space of AE to compress high fidelity data. \textcolor{rev1}{Due to their modular nature, such POD-assisted deep neural network approaches have increased interest in a variety of fluid and solid dynamics applications \citep{kherad2021reduced,cai2021acquisition,ahmed2020long,pawar2019deep,san2018neural,san2019artificial,wang2018model,jacquier2021non,huang2020machine,abadia2022predictive,ooi2021modeling,deng2019time,im2021surrogate}. The main idea utilized to construct the NLPOD model is to use two different reduction strategies. The first reduction strategy is the classical POD approach to generate a latent space. The second reduction strategy is to use an AE, followed by an LSTM architecture to learn the dynamics in the latent space. Hence, the utilization of both reduction strategies significantly decreases the ROM dimension and alleviates the Kolmogorov barrier.}  

%With the advantage of using various tools such as AE, LSTM and ..., ML forms a combined framework of various tools that is able to learn complex and nonlinear patterns existing in the desired physical fields.
% ,,ahmed2021nonlinear

%=============================
%6. More ML papers cited with the compunation of AE and LSTM.. You can refer our NLPOD paper.. In a recent work, Shady et al introduced an NLPOD approach, which is a systematic or synergistic combination of POD, autoencoder and lstm technologies. 
%=============================
In this work, \textcolor{rev1}{our main contribution is a construction of a systematic study on the NLPOD approach, a hybrid POD-LSTM surrogate modeling approach recently introduced by \citet{ahmed2021nonlinear}. We extend the NLPOD method for the Rayleigh Bénard convection (RBC) problem to perform a detailed analysis on a variety of configurations with differing degrees of complexity.} This problem introduces Rayleigh number, which controls the irregularity of underlying dynamics. Increasing Rayleigh number makes an increase in Kolmogorov n-width, and offers more challenging tests for model reduction studies. In addition, in our work, we have performed an uncertainly quantification study of the proposed NLPOD approach considering a wide range of hyperparameters such as the learning rate, initialization, optimization algorithm, activation function, number of LSTM blocks, and number of units in LSTM layers.

%=============================
The rest of the paper is structured as follows. We describe governing equations of two-dimensional RBC in Section~\ref{governingEq}. Next, we introduce numerical methods used for generation of high fidelity data in Section~\ref{NumericalScheme}. In Section~\ref{pod}, POD is introduced for the first phase of data compression, then, in Section~\ref{ae}, AE is utilized for the second phase of data compression. Next, we feed the latent space data into the LSTM networks for forecasting, which is explained in Section~\ref{lstm}. Finally, our numerical results are discussed in Section~\ref{results} with concluding remarks drawn in Section~\ref{conclusions}.
%=============================

%%%%%%%%%%%%%%%%%%%%%%%%%%%%%%%%%%%%%%%%%%%%%%%%%%
\section{Governing equations for two-dimensional Bousinessq flow}
\label{governingEq}
%%%%%%%%%%%%%%%%%%%%%%%%%%%%%%%%%%%%%%%%%%%%%%%%%%
Atmospheric and oceanic circulations caused by temperature difference can be modeled with the Boussinesq approximation to capture geophysical waves \citep{majda2003introduction}. The two-dimensional (2D) dimensionless form of Navier-Stokes equations for incompressible flow with the Boussinesq approximation can be written as:
%-----------------------------------
\begin{equation}\label{eq:continuity}
    \frac{\partial u}{\partial x}+\frac{\partial v}{\partial y}=0,
\end{equation}
%-----------------------------------
\begin{equation}\label{eq:momentumX}
    \frac{\partial u}{\partial t}+u\frac{\partial u}{\partial x}+v\frac{\partial u}{\partial y}=-\frac{\partial p}{\partial x}+\Big(\frac{\text{Pr}}{\text{Ra}}\Big)^{0.5}\Big(\frac{\partial^2 u}{\partial x^2}+\frac{\partial^2 u}{\partial y^2}\Big),
\end{equation}
%-----------------------------------
\begin{equation}\label{eq:momentumY}
    \frac{\partial v}{\partial t}+u\frac{\partial v}{\partial x}+v\frac{\partial v}{\partial y}=-\frac{\partial p}{\partial y}+\Big(\frac{\text{Pr}}{\text{Ra}}\Big)^{0.5}\Big(\frac{\partial^2 v}{\partial x^2}+\frac{\partial^2 v}{\partial y^2}\Big)+\theta,
\end{equation}
%-----------------------------------
\begin{equation}\label{eq:temperature}
    \frac{\partial \theta}{\partial t}+u\frac{\partial \theta}{\partial x}+v\frac{\partial \theta}{\partial y}+\Big(\frac{1}{\text{Ra}\,\text{Pr}}\Big)^{0.5}\Big(\frac{\partial^2 \theta}{\partial x^2}+\frac{\partial^2 \theta}{\partial y^2}\Big),
\end{equation}
%-----------------------------------
where $\theta$ and $p$ are the temperature and pressure, respectively. Since the flow is 2D, the velocity vector field $\pmb{u}=(u,v)$ has horizontal and vertical components. Prandtl number (Pr), the ratio of kinematic viscosity to the thermal diffusivity, and Rayleigh number (Ra), the balance between the gravitational forces and viscous damping, are two dimensionless parameters.

%To avoid checkerboard numerical instability for pressure, we remove pressure by converting the above equations to vorticity-streamfunction system by taking the curl operator of the Eq.~\ref{eq:momentumX} and Eq.~\ref{eq:momentumY} and using the definition of the vorticity vector $\vec{\omega}$ as $\vec{\omega}=\nabla\times\pmb{u}$. 
By changing the aforementioned equations to a vorticity-streamfunction system utilizing the curl operator of the equations for Eq.~\ref{eq:momentumX} and Eq.~\ref{eq:momentumY}, and the definition of the vorticity vector $\boldsymbol \omega$ as $\boldsymbol \omega=\nabla\times\pmb{u}$, we may avoid the numerical instability that results from pressure checkerboarding.
Dealing with a 2D flow problem, we only consider the z-component of the vorticity vector, denoted as $\omega$ henceforth. The relationship between velocity components and streamfunction $\psi$, which satisfies continuity equation, is defined as follows:
%-----------------------------------
\begin{equation}\label{eq:vel}
u=\frac{\partial \psi}{\partial y}, \;\;\;\;\; v=-\frac{\partial \psi}{\partial x}.
\end{equation}
%-----------------------------------
Taking derivative of Eq.~\ref{eq:vel} yields Eq.~\ref{eq:continuitySW} to link vorticity with streamfunction. The following equations represent the \textcolor{rev1}{vorticity-streamfunction} formulation of the fluid flow:
%-----------------------------------
\begin{equation}\label{eq:continuitySW}
\frac{\partial^2 \psi}{\partial x^2}+\frac{\partial^2 \psi}{\partial y^2}=-\omega,
\end{equation}
%-----------------------------------
\begin{equation}\label{eq:momentumSW}
    \frac{\partial \omega}{\partial t}+\frac{\partial \psi}{\partial y}\frac{\partial \omega}{\partial x}-\frac{\partial \psi}{\partial x}\frac{\partial \omega}{\partial y}=-\frac{\partial p}{\partial x}+\Big(\frac{\text{Pr}}{\text{Ra}}\Big)^{0.5}\Big(\frac{\partial^2 \omega}{\partial x^2}+\frac{\partial^2 \omega}{\partial y^2}\Big)+\frac{\partial \theta}{\partial x},
\end{equation}
%-----------------------------------
\begin{equation}\label{eq:temperatureSW}
    \frac{\partial \theta}{\partial t}+\frac{\partial \psi}{\partial y}\frac{\partial \theta}{\partial x}-\frac{\partial \psi}{\partial x}\frac{\partial \theta}{\partial y}+\Big(\frac{1}{\text{Ra}\,\text{Pr}}\Big)^{0.5}\Big(\frac{\partial^2 \theta}{\partial x^2}+\frac{\partial^2 \theta}{\partial y^2}\Big).
\end{equation}
%-----------------------------------

\section{Methodology}
\label{Methodology}

%=========================================
\subsection{Numerical methods}
\label{NumericalScheme}
%=========================================
We briefly describe the numerical methods used to acquire full order model (FOM) data for this study. We use the Pad\'{e} scheme that has been explained in \cite{lele1992compact}. A general Pad\'{e} scheme for the first derivative is:
%-----------------------------------
\begin{equation}\label{eq:fd1}
\alpha f^{\prime}_{i-1}+f^{\prime}_{i}+\alpha f^{\prime}_{i+1}=a\frac{f_{i+1}-f_{i-1}}{2h}+b\frac{f_{i+2}-f_{i-2}}{4h},
\end{equation}
%-----------------------------------
where $a=\frac{2}{3}(\alpha+2)$ and $b=\frac{1}{3}(4\alpha-1)$. Here, $h$ represents the spacing of the uniform spatial grid and the subscript $i$ is an index for nodes. We choose $\alpha=\frac{1}{4}$ to make $b=0$. Therefore, we have a three-point stencil scheme to provide a fourth-order truncation error as follows:
%-----------------------------------
\begin{equation}\label{eq:fd2}
\frac{1}{4} f^{\prime}_{i-1}+f^{\prime}_{i}+\frac{1}{4} f^{\prime}_{i+1}=\frac{3}{4h}(f_{i+1}-f_{i-1}).
\end{equation}
%-----------------------------------
In addition, the general Pad\'{e} scheme for the second derivative is:
%-----------------------------------
\begin{equation}\label{eq:fdd1}
\alpha f^{\prime\prime}_{i-1}+f^{\prime\prime}_{i}+\alpha f^{\prime\prime}_{i+1}=a\frac{f_{i+1}-2f_{i}+f_{i-1}}{2h^{2}}+b\frac{f_{i+2}-2f_{i}+f_{i-2}}{4h^{2}},
\end{equation}
%-----------------------------------
where $a=\frac{4}{3}(1-\alpha)$ and $b=\frac{1}{3}(10\alpha-1)$. Similarly, we set $\alpha=\frac{1}{10}$ to make $b=0$ and have a three-point stencil with the fourth-order truncation error for the second derivative that is given as:
%-----------------------------------
\begin{equation}\label{eq:fdd2}
\frac{1}{10} f^{\prime\prime}_{i-1}+f^{\prime\prime}_{i}+\frac{1}{10} f^{\prime\prime}_{i+1}=\frac{6}{5h^{2}}(f_{i+1}-2f_{i}+f_{i-1}).
\end{equation}
%-----------------------------------
A general high-order Pad\'{e} method is written for the left boundary condition of the first derivative as follows:
%-----------------------------------
\begin{equation}\label{eq:bcfd1}
f^{\prime}_{0}+\alpha f^{\prime}_{1}=\frac{1}{h}(af_{0}+bf_{1}+cf_{2}+df_{3}),
\end{equation}
%-----------------------------------
where $a=-\frac{11+2\alpha}{6}$, $b=\frac{6-\alpha}{2}$, $c=\frac{2\alpha-3}{2}$, and $d=\frac{2-\alpha}{6}$.
We select $\alpha=2$ to make $d=0$ and have a three-point stencil for the first derivative at the boundary, which is given as:
%-----------------------------------
\begin{equation}\label{eq:bcfd2}
f^{\prime}_{0}+2f^{\prime}_{1}=\frac{1}{2h}(-5f_{0}+4f_{1}+f_{2}).
\end{equation}
%-----------------------------------
A general high-order Pad\'{e} method that is written for the left boundary condition of the second derivative follows:
%-----------------------------------
\begin{equation}\label{eq:bcfdd1}
f^{\prime\prime}_{0}+\alpha f^{\prime\prime}_{1}=\frac{1}{h^{2}}(af_{0}+bf_{1}+cf_{2}+df_{3}+ef_{4}),
\end{equation}
%-----------------------------------
where $a=\frac{11\alpha+35}{12}$, $b=-\frac{5\alpha+26}{3}$, $c=\frac{\alpha+19}{2}$, $d=\frac{\alpha-14}{3}$, and $e=\frac{11-\alpha}{12}$.
We set $\alpha=11$ to make $e=0$ and have a four-point stencil for the second derivative at the boundary that is given as:
%-----------------------------------
\begin{equation}\label{eq:bcfdd2}
f^{\prime\prime}_{0}+11f^{\prime\prime}_{1}=\frac{1}{h^{2}}(13f_{0}-27f_{1}+15f_{2}-f_{3}).
\end{equation}
%-----------------------------------
%$\text{CFL}=0.9$
%-----------------------------------
The procedure for marching in time demands the solution of Eq.~\ref{eq:continuitySW}, $\nabla^{2}\psi=-\omega$, at each time step. The discrete Poisson equation can be written as follows \citep{san2015novel}:
%-----------------------------------
\begin{equation}\label{eq:poisson1}
\begin{split}
a\psi_{i,j}+b(\psi_{i+1,j}+\psi_{i-1,j})+c(\psi_{i,j+1}+\psi_{i,j-1})+\\
d(\psi_{i+1,j+1}+\psi_{i+1,j-1}+\psi_{i-1,j+1}+\psi_{i-1,j-1})=\\
-(\frac{\Delta x^{2}}{2})(8\omega_{i,j}+\omega_{i+1,j}+\omega_{i-1,j}+\omega_{i,j+1}+\omega_{i,j-1}),
\end{split}
\end{equation}
%-----------------------------------
where $a=-10(1+\mu^{2})$, $b=5-\mu^{2}$, $c=5\mu^{2}-1$, $d=\frac{1+\mu^{2}}{2}$, and $\mu=\frac{\Delta x}{\Delta y}$.
To overcome computational challenge of solving the Poisson equation, we use the fast Fourier transform, which allows us to utilize the Thomas algorithm along $y$ direction:
%-----------------------------------
\begin{equation}\label{eq:FFT1}
\hat{\omega}_{k,j}=\frac{1}{N_{x}} \sum_{i=0}^{N_{x}-1}\omega_{i,j}\exp{(-I\frac{2\pi ki}{N_{x}})},
\end{equation}
%-----------------------------------
where $i=0,1,2,...,N_{x}$, $j=0,1,2,...,N_{y}$, and $I^{2}=-1$. The following system of algebraic equations is obtained by applying the transform to the above nine-point stencil:
%-----------------------------------
\begin{equation}\label{eq:FFT2}
\alpha_{k}\hat{\psi}_{k,j-1}+\beta_{k}\hat{\psi}_{k,j}+\alpha_{k}\hat{\psi}_{k,j+1}=\hat{R}_{k,j},
\end{equation}
%-----------------------------------
where
%-----------------------------------
\begin{equation}\label{eq:FFT3}
\begin{split}
\alpha_{k}=c+2d\cos{(\frac{2\pi k}{N_{x}})},\\
\beta_{k}=a+2b\cos{(\frac{2\pi k}{N_{x}})},\\
\hat{R}=-(\frac{\Delta x^{2}}{2})(\hat{\omega}_{k,j-1}+
(8+2\cos{(\frac{2\pi k}{N_{x}})})\hat{\omega}_{k,j}+\hat{\omega}_{k,j+1}).
\end{split}
\end{equation}
%-----------------------------------
The inverse Fourier transform yields the solution in physical space as follows:
%-----------------------------------
\begin{equation}\label{eq:FFT4}
\psi_{i,j}=\sum_{k=\frac{-N_{x}}{2}}^{\frac{N_{x}}{2}-1}
\hat{\psi}_{k,j}\exp{(I\frac{2\pi ki}{N_{x}})}.
\end{equation}
%-----------------------------------
While we assign periodic boundary conditions along $x$ direction, we use impermeability $\frac{\partial \psi}{\partial n}=0$ and no-slip condition $\psi=0$ on the top and bottom boundaries to set boundary condition for the vorticity and streamfunction variables. Details are available in \cite{weinan1996essentially} and \cite{briley1971numerical}. Hence, the boundary condition for the vorticity reads as follows:
%-----------------------------------
\begin{equation}\label{eq:bcOmega}
\omega_{0}=\frac{1}{h^{2}}(-6\psi_{1}+\frac{3}{2}\psi_{2}-\frac{2}{9}\psi_{3}). %\frac{85}{18}\psi_{0}
\end{equation}
%-----------------------------------
After spatial discretization, we need to solve the semi-discrete ordinary differential equations (ODEs) along time with the third order Runge-Kutta method. In the following, we represent a semi-discrete system of ODEs:
%-----------------------------------
\begin{equation}\label{eq:RK30}
%y^{\prime}=\text{f}(y)
z^{\prime}=\mathfrak{R}(z),
\end{equation}
where $\mathfrak{R}(\cdot)$ denotes all the remaining terms with spatial derivatives.
%-----------------------------------
The third-order Runge-Kutta method is as follows:
%-----------------------------------
\begin{equation}\label{eq:RK31}
z^{(1)}=z^{n}+\Delta t\,\mathfrak{R}(z^{n}),
\end{equation}
%-----------------------------------
\begin{equation}\label{eq:RK32}
z^{(2)}=\frac{3}{4}y^{n}+\frac{1}{4}z^{(1)}+\frac{1}{4}\Delta t\,\mathfrak{R}(z^{(1)}),
\end{equation}
%-----------------------------------
\begin{equation}\label{eq:RK33}
z^{n+1}=\frac{1}{3}y^{n}+\frac{2}{3}z^{(2)}+\frac{2}{3}\Delta t\,\mathfrak{R}(z^{(2)}).
\end{equation}
%-----------------------------------

%=========================================
\subsection{Proper orthogonal decomposition}
\label{pod}
%=========================================
The primary goal of POD is to find a set of optimal linear low-dimensional basis that represents the high-dimensional data. The basis set is optimal because the error between projection and training data is \textcolor{rev1}{minimized in the L2 norm.} We form the matrix $\mathbf{A}$ with spatiotemporal temperature data acquired from FOM simulation as follows:
%-----------------------------------
\begin{equation}\label{eq:snapshot_A}
    \mathbf{A} = \bigg[ \boldsymbol{\theta}^{(1)},\boldsymbol{\theta}^{(2)}, \dots, \boldsymbol{\theta}^{(N_s)}\bigg] \in \mathbb{R}^{n \times N_s},
\end{equation}
%-----------------------------------
where $\theta^{(i)}$ refers to temperature field at $i$th time step. $N_s$ is the total number snapshots and $n$ is the number of spatial degrees of freedom. A singular value decomposition (SVD) of the matrix $\mathbf{A}$ can be carried-out as below:
%-----------------------------------
\begin{equation}\label{eq:POD2}
    \mathbf{A} = \mathbf{U} \boldsymbol{\Sigma} \mathbf{V}^*,
\end{equation}
%-----------------------------------
where $\mathbf{U}\in \mathbb{R}^{n \times n}$ is a matrix filled with eigenvectors of the matrix $\mathbf{A}\mathbf{A}^* \in \mathbb{R}^{n \times n}$ and $\mathbf{V}^*\in \mathbb{R}^{N_s \times N_s}$ is a matrix filled with eigenvectors of the matrix $\mathbf{A}^*\mathbf{A}\in \mathbb{R}^{N_s \times N_s}$. The matrix $\boldsymbol{\Sigma}\in \mathbb{R}^{n \times N_s}$ is filled with the sorted square root $\sigma_1 \geq \sigma_2 \geq ... \geq \sigma_{m} \geq 0$ of the eigenvalues of the matrices $\mathbf{A}\mathbf{A}^*$ or $\mathbf{A}^*\mathbf{A}$ where $m=\text{min}(n,N_s)$.

\textcolor{rev1}{Broadly speaking,} taking full SVD is computationally intensive due to the fact that the number of the grid size in a 2D simulation is usually high. Therefore, we utilize the NumPy package ``\texttt{numpy.linalg.svd}" with the option ``\texttt{full\_matrices=False"} to calculate the reduced or economy version of SVD.

Every eigenvalue points out the significance of its eigenfunction for reconstructing the high dimensional data. As a result, we keep only the first $N_R$ eigenvalues and the corresponding eigenfuncitons to reduce the data and have the most accurate estimate of high dimensional data in the low-rank subspace. POD basis functions are the retained eigenfunctions that are obtained by truncating $U$ matrix as below:
%-----------------------------------
\begin{equation}\label{eq:bases}
    \boldsymbol{\Phi} = \bigg[ \boldsymbol{U}^{(1)},\boldsymbol{U}^{(2)}, \dots, \boldsymbol{U}^{(N_R)}\bigg] \in \mathbb{R}^{n \times N_R}.
\end{equation}
%-----------------------------------
To choose the minimum $N_R$, we use the relative information content (RIC) in a given number of the POD modes as follows:
%-----------------------------------
\begin{equation}\label{eq:content}
    \text{RIC} (N_R) = \dfrac{\sum_{k=1}^{N_R} \sigma_k^2}{\sum_{k=1}^{N_s} \sigma_k^2} 
    \times 100 \, \%.
\end{equation}
%-----------------------------------
When the POD bases are constructed, the low-dimensional temporal coefficients can be calculated as below:
%-----------------------------------
\begin{equation}\label{eq:pod_coeffients}
    {\boldsymbol{\alpha}} = \boldsymbol{\Phi}^T \boldsymbol{\theta}.
\end{equation}
%-----------------------------------
Here, $\boldsymbol{\alpha} \in \mathbb{R}^{N_R \times N_s}$ denotes the vector of temporal POD coefficients. The optimal reconstruction of the temperature can be obtained as follows:
%-----------------------------------
\begin{equation}\label{eq:reconstruction}
    \hat{\boldsymbol{\theta}} ~=~ \boldsymbol{\Phi}{\boldsymbol{\alpha}} ~=~ \boldsymbol{\Phi}\boldsymbol{\Phi}^T \boldsymbol{\theta}.
\end{equation}
%-----------------------------------

%=========================================
\subsection{Autoencoder network}
\label{ae}
\begin{figure}[htbp]
\centering
\includegraphics[width=0.75\textwidth]{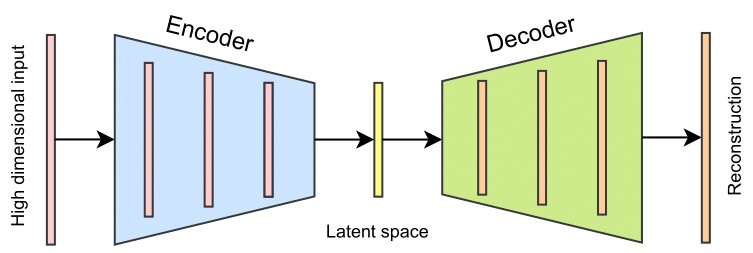}
\caption{Illustration of an autoencoder network which encodes high dimensional inputs to latent space and decodes the latent space to reconstruct inputs.}
\label{fig:ae}
\end{figure}

%=========================================
Autoencoders are unsupervised networks trained to produce the input data along with latent space through learning nonlinear correlations among input features. The encoding layers map the inputs into a latent space with few number of neurons as can be seen in Figure~\ref{fig:ae} and the decoder layers reconstruct the input data given the latent space.
The goal of the AE is to compress the input data into latent space and minimize reconstruction error. An AE with only one hidden layer with a linear activation function acts like the POD. Adding more layers with nonlinear activation functions to the hidden layers serves as nonlinear dimensionality reduction.
%****************************************
%****************************************

%=========================================
\subsection{Long short-term memory network}
\label{lstm}
%=========================================
Recurrent neural networks (RNNs), designed to learn sequential data such as time series, address the stateless issue of classical MLP and CNN by allowing information to persist \citep{45500}. Of particular interest, long short-term memory (LSTM) network, an advanced RNN architecture introduced by \cite{hochreiter1997long}, is able to handle vanishing and exploding gradients problem of RNN, and as a result, it accounts for long-term dependencies. An LSTM network is composed of multiple LSTM blocks consisting of LSTM layers.
The LSTM layers are formed with interacting some LSTM units which are the smallest parts in the LSTM architecture.

%==========================
\subsubsection{LSTM unit}
\begin{figure}[htbp]
\centering
\includegraphics[width=0.75\textwidth]{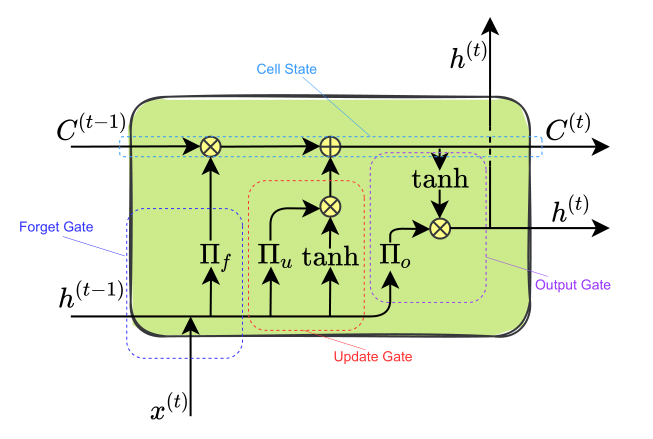}
\caption{The architechture of an LSTM unit.}
\label{fig:LSTMcell}
\end{figure}

%==========================
An LSTM unit shown in Figure~\ref{fig:LSTMcell} consists of a forget gate, an update gate, a cell state, and an output gate.
%****************************************
%****************************************

\paragraph{Forget gate}
%===========
Input and hidden state data enter the forget gate $\mathbf{\Pi}_f$ of the LSTM cell. Forget gate is responsible for storing some information and discarding the rest by using a sigmoid function that is given as
%-----------------------------------
\begin{equation}\label{eq:forget}
    \mathbf{\Pi}_f = \sigma(\mathbf{W}_f [\mathbf{h}^{(t-1)},\mathbf{x}^{(t)}] + \mathbf{b}_f).
\end{equation}
%-----------------------------------
%===========
\paragraph{Update gate}
%===========
The second step is how much information is going to be kept in the cell state. Although $\mathbf{\Pi}_u$ in the update gate has the same mechanism as $\mathbf{\Pi}_f$ in the forget gate, their weights are different. The $\tanh$ normalizes data between -1 and 1 before sending it to $\mathbf{\Pi}_u$ for filtering and feature extraction as
%-----------------------------------
\begin{equation}\label{eq:update}
    \mathbf{\Pi}_u = \sigma(\mathbf{W}_u [\mathbf{h}^{(t-1)},\mathbf{x}^{(t)}] + \mathbf{b}_u),
\end{equation}
%-----------------------------------
\begin{equation}\label{eq:update2}
\tilde{\mathbf{C}}^{(t)} = \text{tanh}(\mathbf{W}_c [\mathbf{h}^{(t-1)},\mathbf{x}^{(t)}] + \mathbf{b}_c).
\end{equation}
%-----------------------------------
%===========
\paragraph{Cell state}
%===========
Insignificant information of the previous cell state entering into the LSTM unit is forgotten through the forget gate. Then, the important information of the cell state is updated through the update gate. As a result, forget and update gates extract features of the cell state that must be remembered. Cell state equation is as follows:
%-----------------------------------
\begin{equation}\label{eq:cellstate}
    \mathbf{C}^{(t)} = \mathbf{\Pi}_f \odot \mathbf{C}^{(t-1)} + \mathbf{\Pi}_u \odot \tilde{\mathbf{C}}^{(t)},
\end{equation}
%-----------------------------------
where $\mathbf{C}^{(t)}$ shows current state of the cell. The gates regulate the LSTM unit to be able to remove or add information to the cell state, carefully regulated by structures called gates.
%===========
\paragraph{Output gate}
%===========
Finally, the cell state for the next time step scaled with a $\tanh$ and filtered with $\mathbf{\Pi}_o$ makes the output. The output provides information that can be utilized for feeding the LSTM unit for the next time step and for the same time step in the next LSTM layer. The equations of the output gate and the next hidden state forecasting are given by:
%-----------------------------------
\begin{equation}\label{eq:outputgate}
    \mathbf{\Pi}_o = \sigma(\mathbf{W}_o [\mathbf{h}^{(t-1)},\mathbf{x}^{(t)}] + \mathbf{b}_o),
\end{equation}
%-----------------------------------
\begin{equation}\label{eq:output}
    \mathbf{h}^{(t)} = \mathbf{\Pi}_o \odot \text{tanh} (\mathbf{C}^{(t)}).
\end{equation}
%-----------------------------------
%****************************************
\begin{figure}[htbp]
\centering
\includegraphics[width=0.75\textwidth]{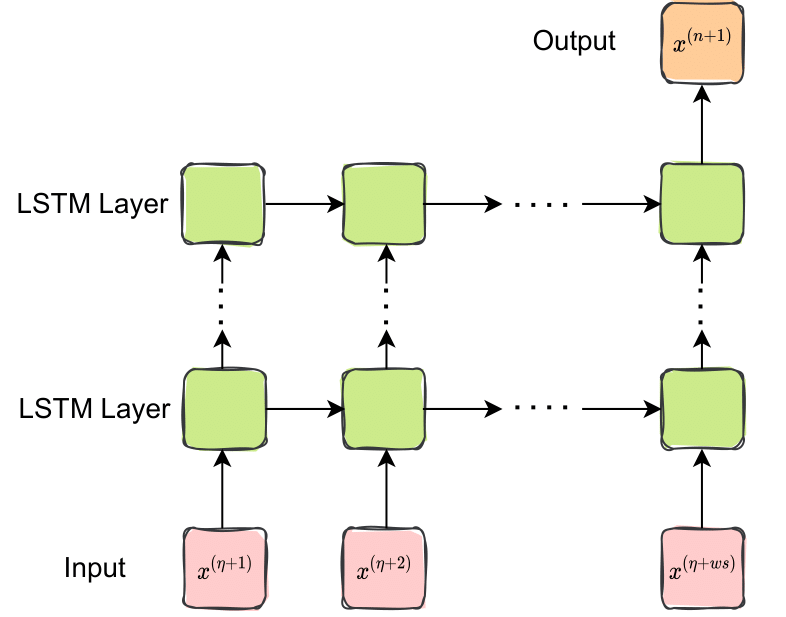}
\caption{Input data $x$ is in red squares, output is in the orange square, and green squares represent hidden layers. The squares from left to right are in the same layer but in different times. The squares from below to up are at the same time but in different LSTM layers.}
\label{fig:LSTMlayers}
\end{figure}
%****************************************
%==========================
\subsubsection{LSTM layers}
%==========================
The set of Eqs.~\ref{eq:forget}--\ref{eq:output} for the LSTM unit has the ability of predicting future state $x^{(n+1)}$ given $x^{(\eta+1)}$, $x^{(\eta+2)}$, $\dots$, $x^{(\eta+ws)}$, where  $\eta=n-ws$. The $ws$ is referred to as the window size determining how many of previous time steps of the temporal information are required to accurately predict the future state of the system. As it can be seen in Figure~\ref{fig:LSTMlayers}, an LSTM network is built with many LSTM units in horizontal lines to create LSTM layers for complexity of the model and in vertical lines to keep information corresponding to a specific time step. Input data whose length is as same as the length of window size is fed to the first layer to predict the hidden state through a many-to-many layer. %The hidden state is passed on to the next layer for similar process to be done. 
\textcolor{rev1}{The following layer receives the hidden state and performs a similar operation.} Finally, last layer, which is a many-to-one layer, forecasts the future state.
%==========================
\subsubsection{LSTM blocks}
\begin{figure}[htbp]
\centering
\includegraphics[width=0.5 \textwidth]{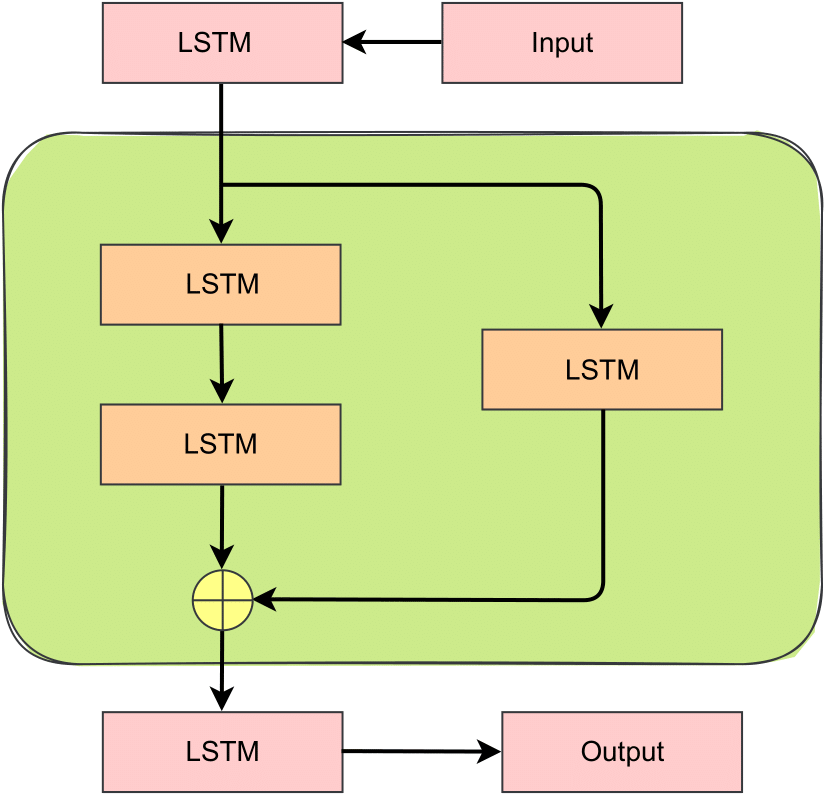}
\caption{Building an LSTM network with LSTM blocks rather than stacking layers consecutively.}
\label{fig:LSTMblock}
\end{figure}

%==========================
Instead of building an LSTM network by adding layers consecutively in series, we make the network with LSTM blocks. Figure~\ref{fig:LSTMblock} shows a network with one block, but we can connect multiple blocks to create a more complex LSTM network.
%****************************************
%****************************************

%=========================================
\subsection{Nonlinear proper orthogonal decomposition}
\label{nlpod}
%=========================================
%Convection-dominated Rayleigh–Bénard flow with high Ra numbers
High dimensional data cannot be accurately reconstructed with limited number of POD modes. Consequently, we choose higher \textcolor{rev1}{number} of POD modes or RIC and remove further uncorrelated data with utilizing autoencoder technology to consider under-resolved flow and to reduce projection error - the error between true projection and NIROM.
In this regard, POD temporal coefficients are fed into multi-layer perceptron AE network to learn much lower dimensional latent space than the number of POD modes. Then, the latent space data is used as input data for LSTM network for time evolution of the low-rank space.
In order to reconstruct high dimensional data in the desired time, first we decode the latent space information after LSTM forecasting, and then, we multiply decoded data or POD temporal coefficients to basis functions.

%========================================================
\section{Numerical results}
\label{results}

%========================================================
We employ the NLPOD framework to forecast temperature in the 2D Rayleigh–Bénard convection flow.
In order to simulate the flow field and acquire the FOM data, we first perform numerical simulations for the $\text{Ra}=2\times10^{6}$, $\text{Ra}=9\times10^{6}$, and $\text{Ra}=1\times10^{7}$ cases on a computational domain covered by $512 \times 256$ nodes.
We collect the temperature field as the FOM data only after the initial transient region.
After taking SVD from $131072$ nodes and keeping $99\%$ of the content, we get $N_R=8,\;28,\;45$ number of POD modes for the $\text{Ra}=2\times10^{6},\;\text{Ra}=9\times10^{6},\;\text{Ra}=1\times10^{7}$, respectively.
We encode the POD modes to reduce the dimension to only $N_r=4$ time series before feeding it to the LSTM network for learning time dependencies.

Training sets for all cases in this paper are from $t=0$ to $t=175$ s, which has colored background in our resulting figures.
Out of sample data is illustrated with white background. We apply the SVD only on the training sets to get the basis functions and employ those basis functions to reconstruct extrapolated temperature field.
We show results only starting from $t=150$ s to remove redundancy.
For the rest of this study, we depict the FOM, \textcolor{rev1}{true projection (TP)}, and mean values as a solid black line, dash-dot green line, and dashed red line, respectively.
The forecasted solution with the AE network is shown by the solid blue line.

In order to thoroughly investigate LSTM performance combined with NLPOD, we train 64 models with different hyperparameters \textcolor{rev1}{(see Table~\ref{table:hyper})}, such as learning rate, activation function, optimizer, initialization, number of LSTM units, and number of LSTM blocks. \textcolor{rev1}{Our analysis also provides the mean $\mu$ and the two standard deviation (SD) bounds $\mu \mp 2\times \text{SD}$.} %More specifically, Figures \ref{fig:fig4},~\ref{fig:fig9},~and \ref{fig:fig14} plot reconstructed temperature on four locations $(1.6,0.1)$, $(1.6,0.15)$, $(2,0.1)$, and $(2,0.15)$, while Figures \ref{fig:fig5},~\ref{fig:fig10},~and \ref{fig:fig15} illustrate the temperature field at $t=195$ s and $t=198$ s.

\setlength{\tabcolsep}{5pt} % Default value: 6pt
\renewcommand{\arraystretch}{1.5} % Default value: 1
\begin{table*}[htbp!]
\caption{\textcolor{rev1}{Hyperparameters used in long short-term memory (LSTM) networks. We note that our analysis involves 64 different LSTM models.} } \vspace{5pt}
\centering
\fontsize{6.5}{12}
\begin{tabular}{ p{5cm} p{5cm}  }
\hline
Parameters & Range of consideration \\
\hline
Learning rate & [0.001, 0.0001] \\
Activation function  & [tanh, relu] \\
Optimizer  & [adam, rmsprop] \\
Initialization  & [uniform, Glorot]\\
Number of LSTM units & [10, 20] \\
Number of LSTM block & [2, 3] \\
\hline
\end{tabular}
\label{table:hyper}
\end{table*}

%focus on the performance of the proposed models both in interpolatory and extrapolatory intervals/windows.

%========================================================
\subsection{$Ra=2\times10^{6}$Case}
\begin{figure}[htbp]
	\centering
	\includegraphics[scale=0.75]{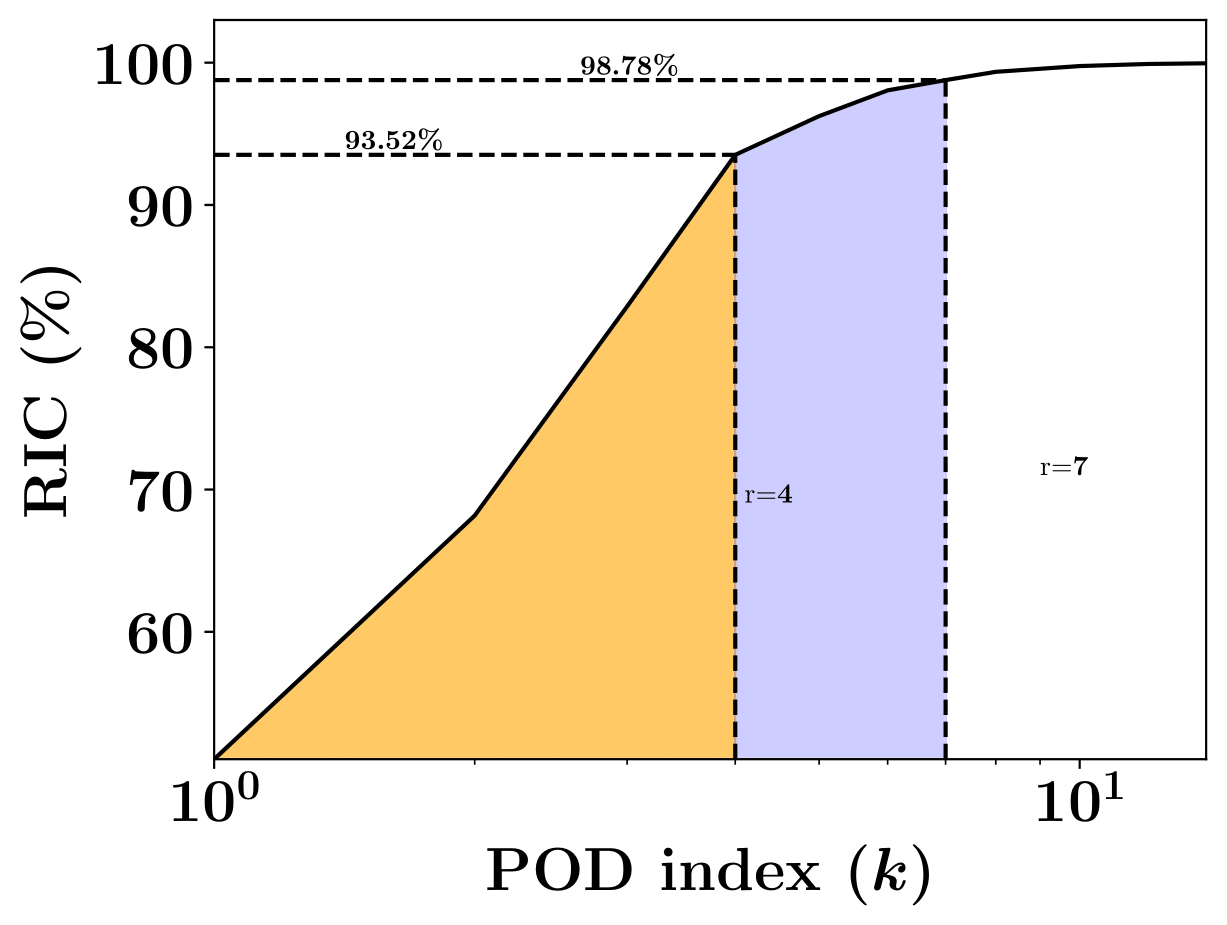}
	\caption{The relative information content based on number of retained modes for $\text{Ra}=2\times 10^{6}$. While the leading $4$ POD modes capture $93.52\%$, the first $8$ POD modes are able to capture
    more that $99\%$ of the relative information content of the
    high dimensional data. As a result, the suggested AE is constructed
    to learn a latent space from $8$ POD coefficients.}
	\label{fig:content1}
\end{figure}

%========================================================
%--------------------------------
For almost periodic dynamics case with $\text{Ra}=2\times 10^6$, Figure \ref{fig:content1} shows that $4$ POD modes are able to achieve $93.52\%$ of the RIC and $8$ modes are needed to capture $99\%$ of the content.

%****************************************
\begin{figure}
	\centering
	\includegraphics[scale=0.3]{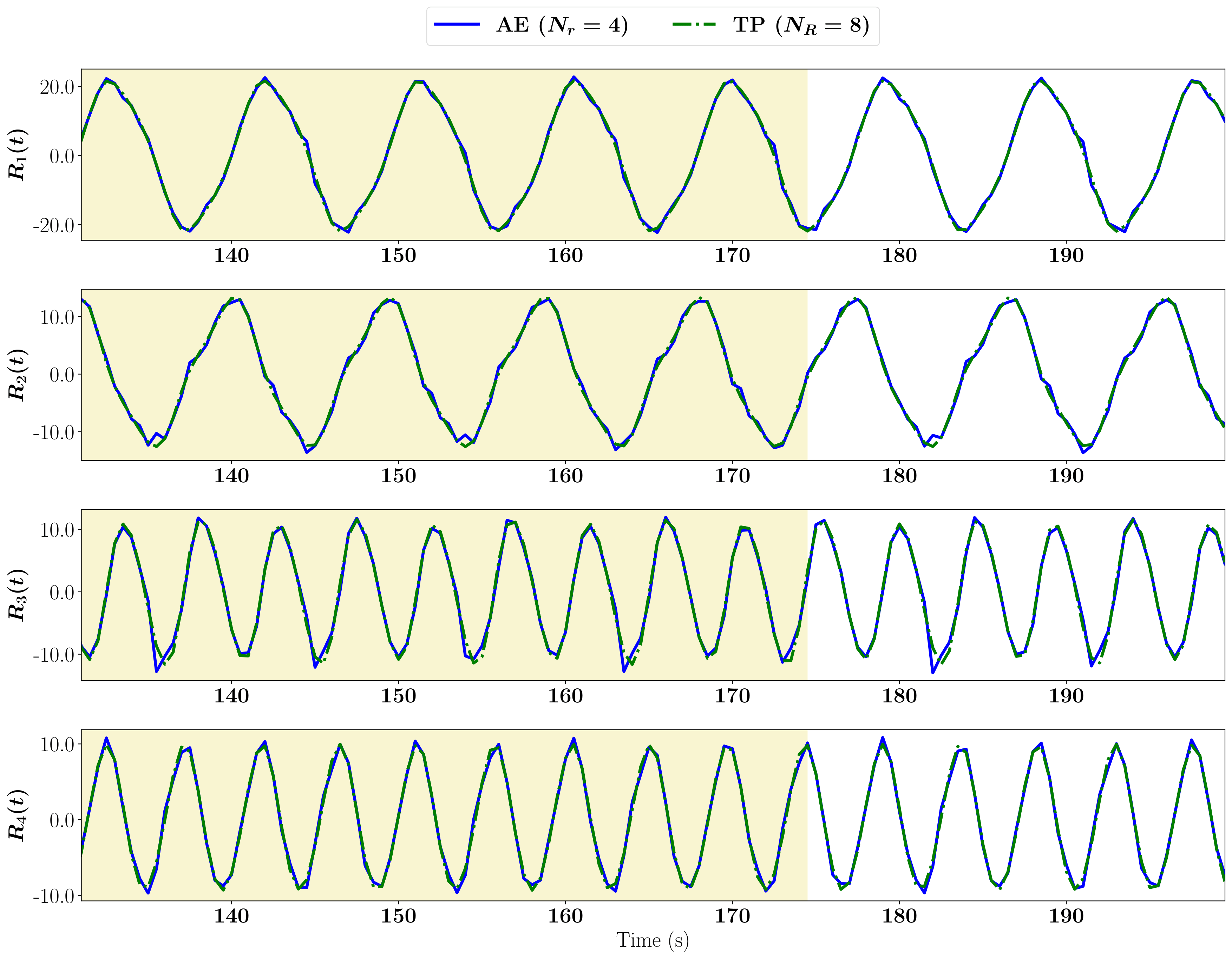}
	\caption{Evolution of the first four POD modal coefﬁcients 
	%$\alpha_1,\;\alpha_2,\;\alpha_3,\;\alpha_4$
	for $\text{Ra}=2\times 10^{6}$.
	The training set for the AE network is from $t = 0$ to $t=175$ s.}
	\label{fig:fig2}
\end{figure}

%****************************************
Figure \ref{fig:fig2} shows the time evolution of the first $4$ POD modes for $\text{Ra}=2\times 10^{6}$.
It can be seen that the AE reconstruction is \textcolor{rev1}{very accurate} for an almost time periodic dynamical system. Therefore, $N_r=4$ is sufficient for this case to reconstruct the POD modes and finally the temperature from the latent space.

%****************************************
\begin{figure}[htbp]
	\centering
	\includegraphics[scale=0.3]{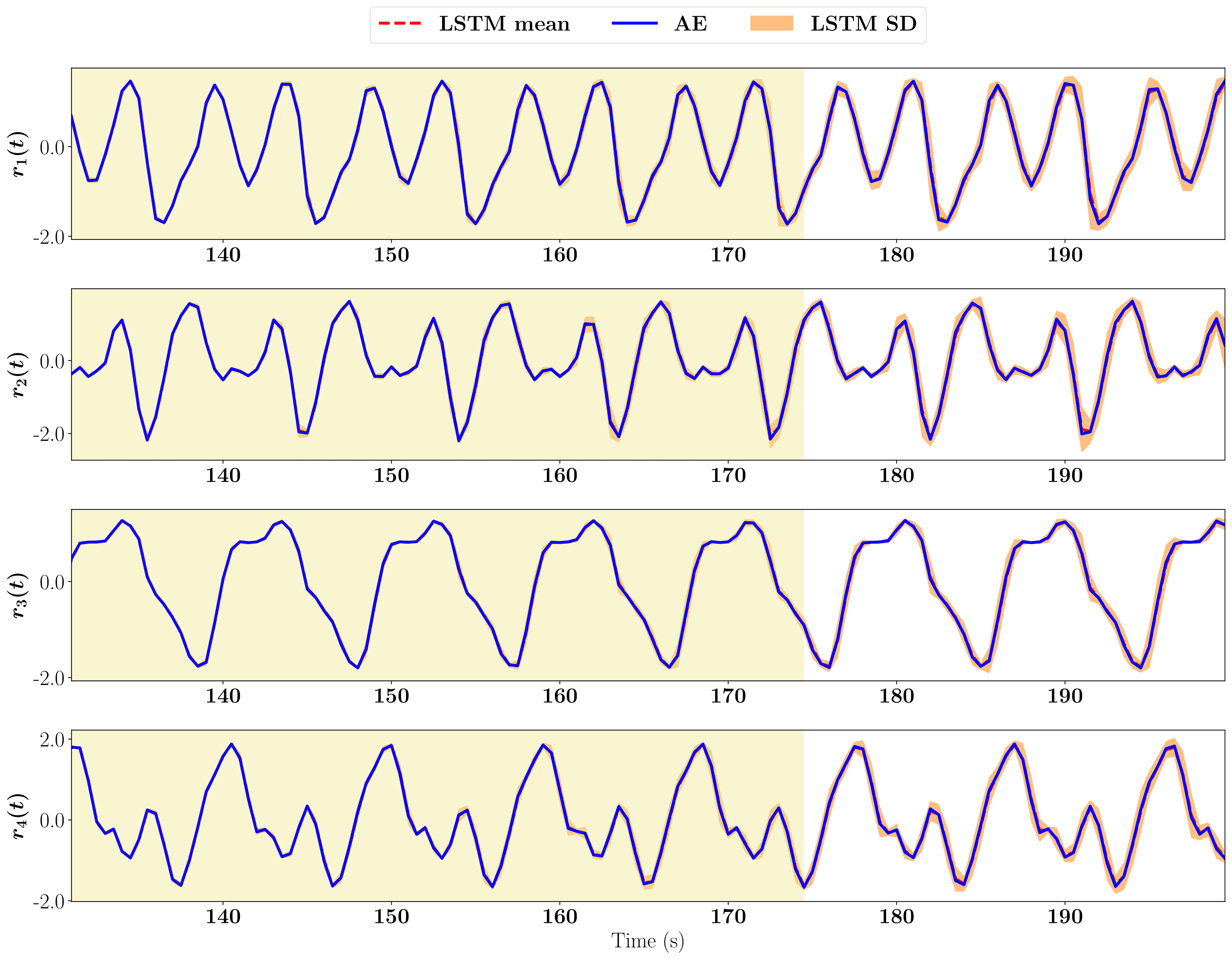}
	\caption{Evolution of autoencoder modes in latent space with the LSTM for $\text{Ra}=2\times 10^{6}$.
	The training set for the LSTM network is from $t = 0$ to $t=175$ s.}
	\label{fig:fig3}
\end{figure}

%****************************************
Figure \ref{fig:fig3} presents the time evolution of four time series in the latent space of the AE network using the data for $\text{Ra}=2\times 10^6$.
The purpose of this plot is to reveal how LSTM performs on predicting of the almost periodic time series.
As it is illustrated in Figure~\ref{fig:fig3}, the $\text{SD}$ bars are quite close to the mean value, meaning we do not need to spend resources for tuning the network to get accurate prediction as long as hyperparameters are in a reasonable range.

%****************************************
\begin{figure}[htbp]
	\centering
	\includegraphics[scale=0.3]{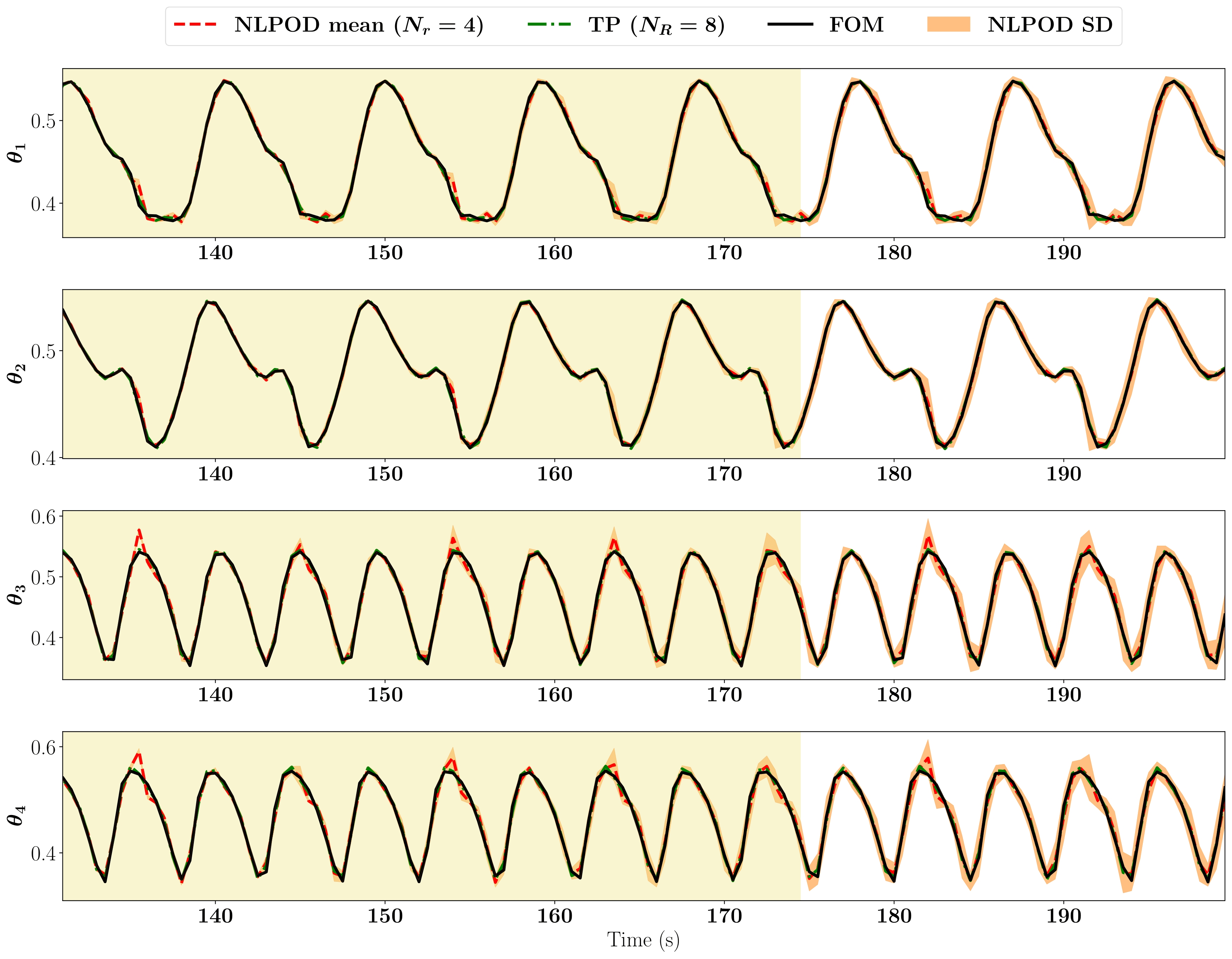}
	\caption{Temperature evolution in physical space for
	$\text{Ra}=2\times 10^{6}$ at four locations.}
	\label{fig:fig4}
\end{figure}

\begin{figure}[htbp]
	\centering
	\includegraphics[scale=0.5]{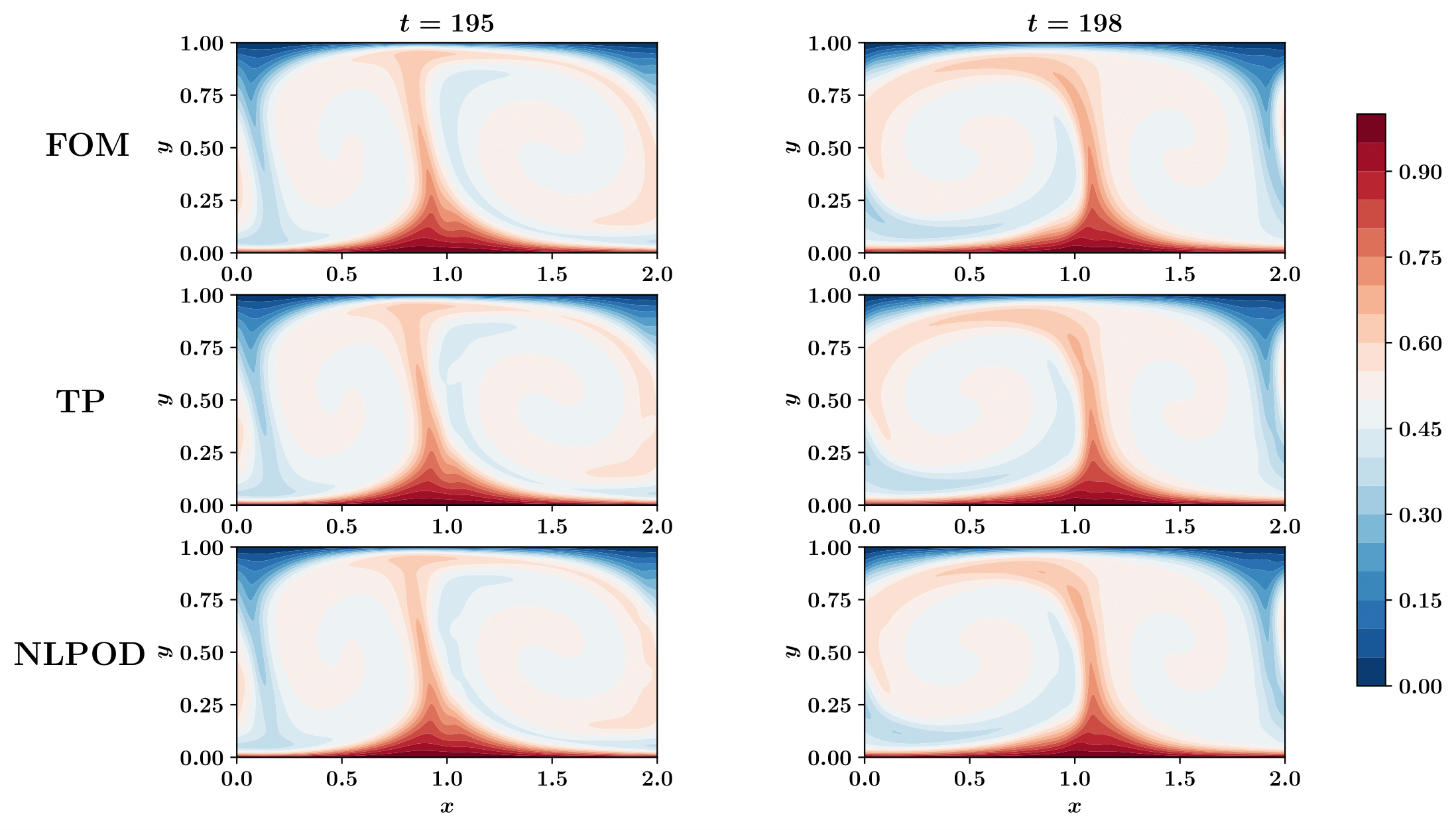}
	\caption{A comparison of temperature fields for $\text{Ra}=2\times 10^{6}$ at two different times.}
	\label{fig:fig5}
\end{figure}

%****************************************
Figure \ref{fig:fig4} plots reconstructed temperature on the four points. As we expect based on the performance of AE and LSTM, farecasted temperature with NLPOD is able to follow the TP curve which is quite close to the FOM solution.
As a result, NLPOD is quite accurate for flows with almost periodic dynamics. This can also be seen from Figure~\ref{fig:fig5} when we compare temperature fields for $\text{Ra}=2\times 10^{6}$ at two different times.

%****************************************
%****************************************
%The contours for the temperature field are reconstructed from NLPOD and TP frameworks and are shown in figure \ref{fig:fig5} at $t=195$ and $t=198$ s to compare them with the FOM data.

%****************************************
%****************************************
%--------------------------------
%=========================================
\subsection{$Ra=9\times10^{6}$Case}
\begin{figure}[htbp]
	\centering
	\includegraphics[scale=0.75]{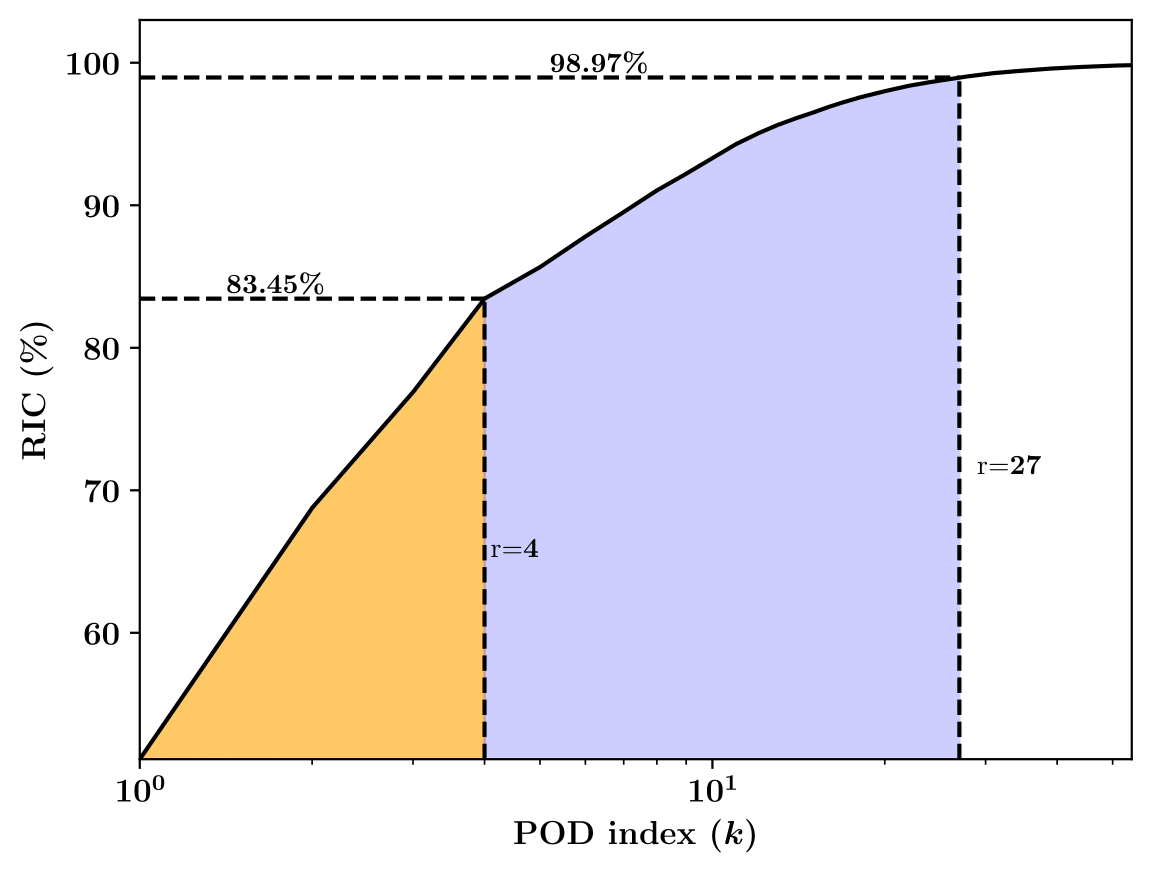}
	\caption{The relative information content based on number of retained modes for $\text{Ra}=9\times 10^{6}$. While the leading $4$ POD modes capture $83.45\%$, the first $28$ POD modes are able to capture
    more that $99\%$ of the relative information content of the
    high dimensional data. As a result, the suggested AE is constructed
    to learn a latent space from $28$ POD coefficients.}
	\label{fig:content2}
\end{figure}

%=========================================
%--------------------------------
For quasi-periodic dynamics with $\text{Ra}=9\times 10^6$, Figure \ref{fig:content2} shows that $4$ POD modes have the ability to acheive $83.45\%$ of the RIC and $28$ modes are needed to capture more than $99\%$ of the content.

%****************************************
\begin{figure}[htbp]
	\centering
	\includegraphics[scale=0.3]{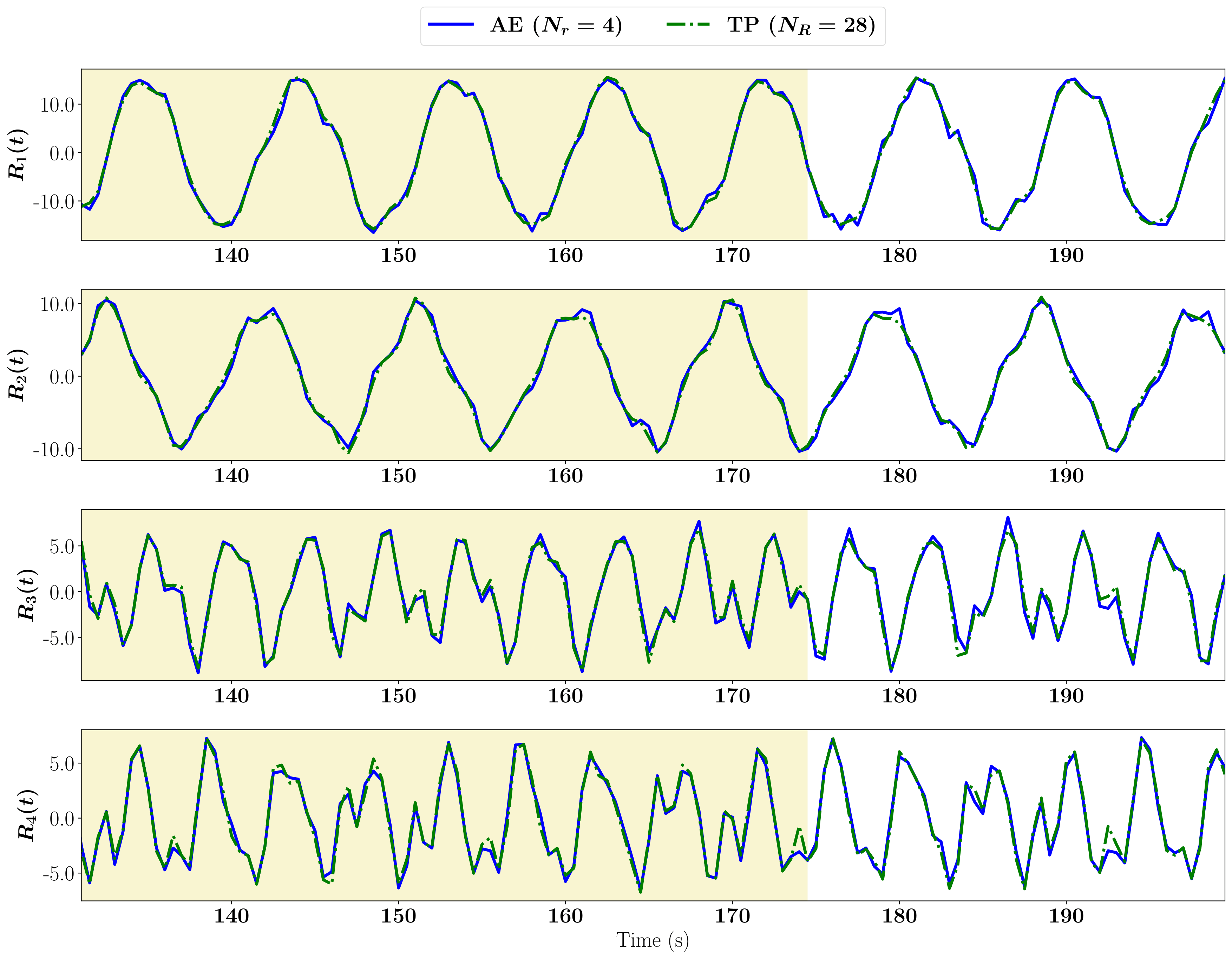}
	\caption{Variation of first four POD temporal coefﬁcients 
	%$\alpha_1,\;\alpha_2,\;\alpha_3,\;\alpha_4$
	for $\text{Ra}=9\times 10^{6}$.
	The training set for the AE network is from $t = 0$ to $t=175$ s.}
	\label{fig:fig7}
\end{figure}

%****************************************
Figure \ref{fig:fig7} illustrates the time evolution of the first $4$ POD modes for $\text{Ra}=9\times 10^{6}$.
It can still be seen that the AE reconstruction is pretty close for a quasi-periodic dynamical system, and as a result, the AE is not a bottleneck for this case with $\text{Ra}=9\times 10^{6}$. 

%****************************************
\begin{figure}[htbp]
	\centering
	\includegraphics[scale=0.3]{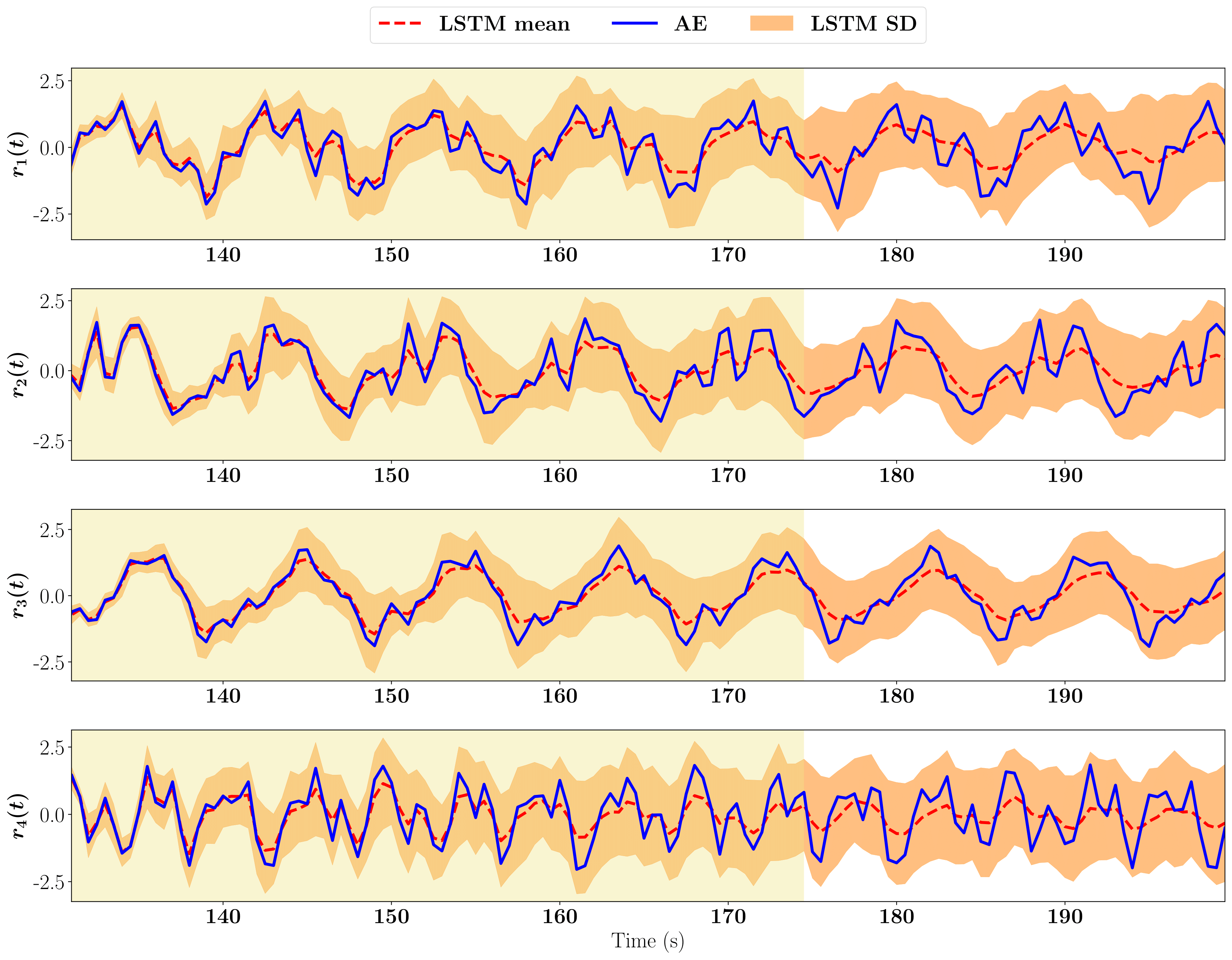}
	\caption{Evolution of autoencoder modes in latent space with the LSTM for $\text{Ra}=9\times 10^{6}$.
	The training set for the LSTM network is from $t = 0$ to $t=175$ s.}
	\label{fig:fig8}
\end{figure}

%****************************************
Figure \ref{fig:fig8} illustrates the time evolution in the latent space for four neurons of the AE network utilizing the data for $\text{Ra}=9\times 10^6$.
Our aim of drawing this figure is to demonstrate the LSTM performance is accurate for quasi-periodic dynamics in the latent space of AE with some hyperparameter tuning. As it is shown in Figure \ref{fig:fig8} the $\text{SD}$ bars are close to the mean value in the interpolatory region, but they have a little more width in the extrapolatory region.

%****************************************
\begin{figure}[htbp]
	\centering
	\includegraphics[scale=0.3]{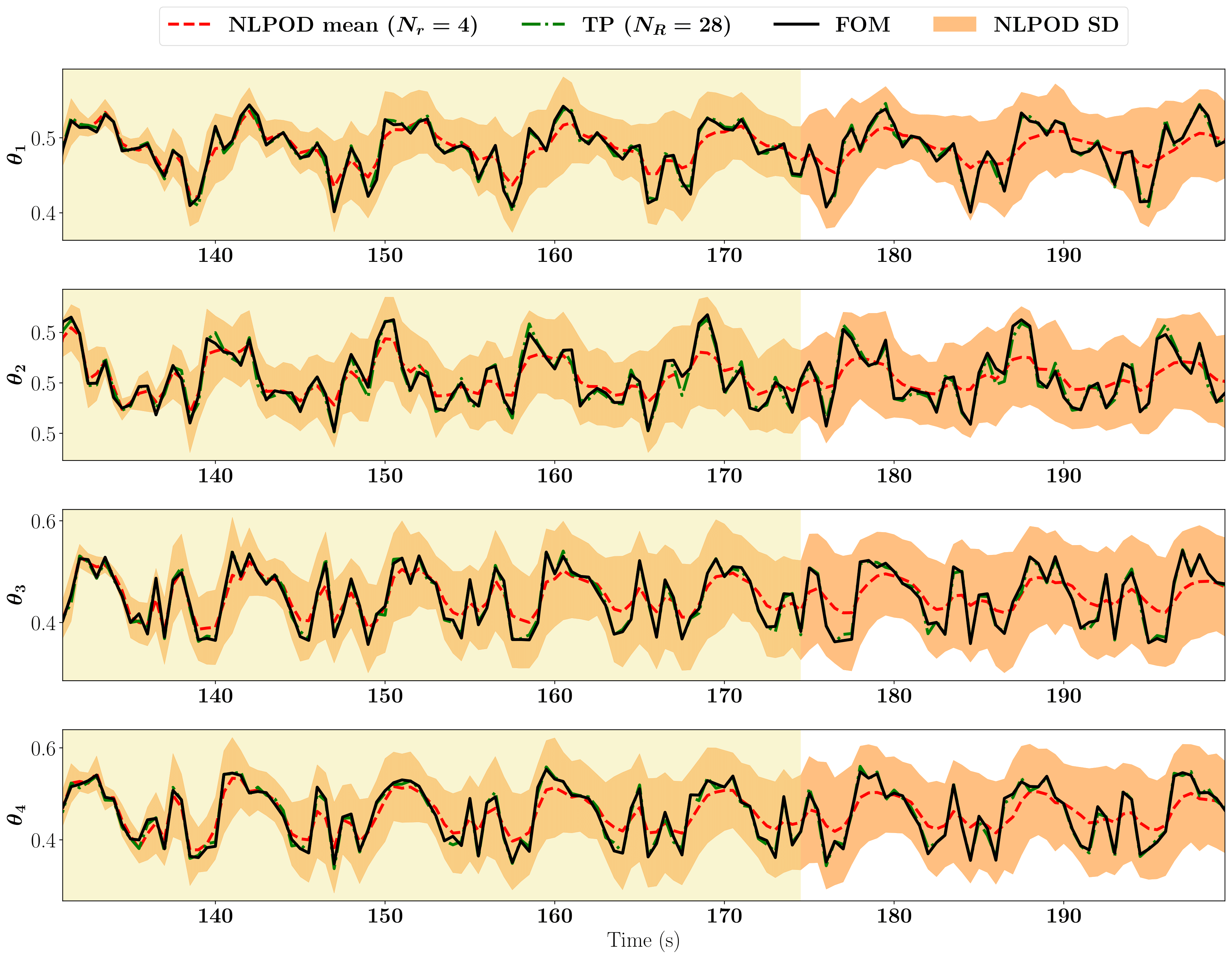}
	\caption{Temperature evolution in physical space for
	$\text{Ra}=9\times 10^{6}$ at four locations.}
	\label{fig:fig9}
\end{figure}

\begin{figure}[htbp]
	\centering
	\includegraphics[scale=0.5]{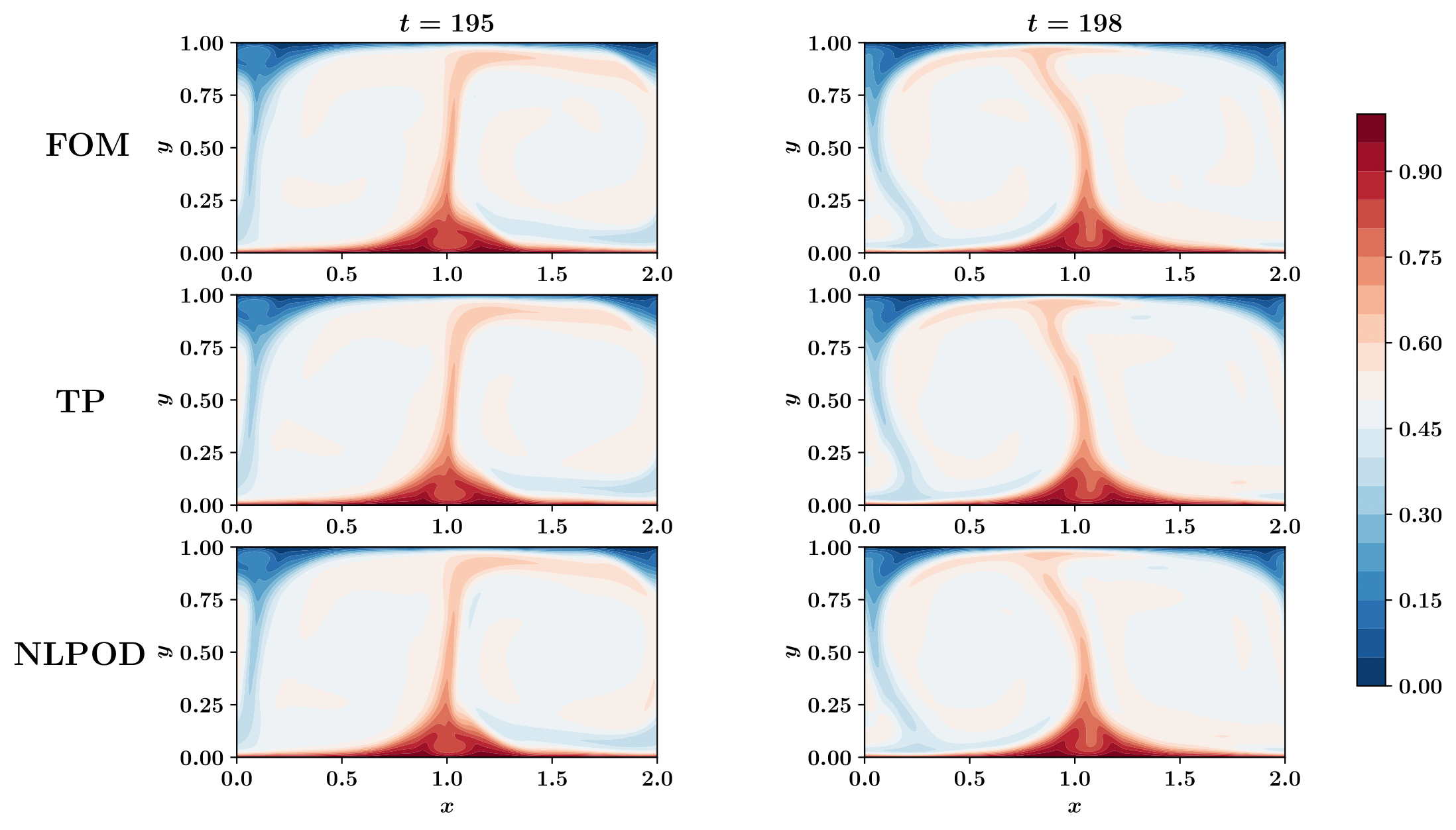}
	\caption{A comparison of temperature fields for $\text{Ra}=9\times 10^{6}$ at two different times.}
	\label{fig:fig10}
\end{figure}

%****************************************
Figure \ref{fig:fig9} compares the reconstructed temperature field of FOM, TP, and NLPOD at four locations.
According to Figure~\ref{fig:fig9}, TP is close to FOM data because we retain $99\%$ of content by keeping $28$ POD modes. This is also verified in Figure~\ref{fig:fig10} for the reconstruction of the temperature fields at two different times.

%****************************************
%****************************************
%****************************************
%--------------------------------
%=========================================
\subsection{$Ra=1\times10^{7}$Case}
\begin{figure}[htbp]
	\centering
	\includegraphics[scale=0.75]{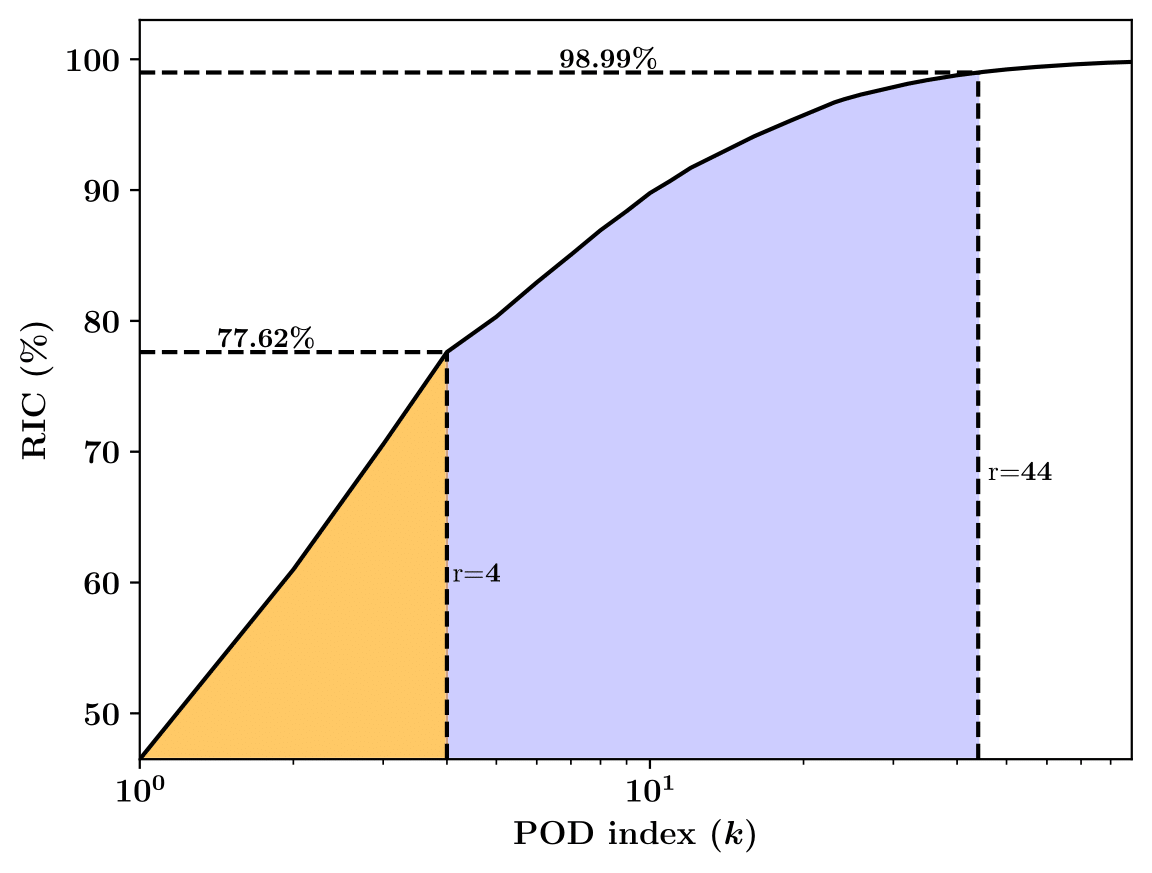}
	\caption{The relative information content based on number of retained modes for $\text{Ra}=1\times 10^{7}$. While the leading $4$ POD modes capture $77.62\%$, the first $45$ POD modes are able to capture
    more that $99\%$ of the relative information content of the
    high dimensional data. As a result, the suggested AE is constructed
    to learn a latent space from $45$ POD coefficients.}
	\label{fig:content3}
\end{figure}

%=========================================
%--------------------------------
For almost chaotic dynamics or irregular dynamics with $\text{Ra}=1\times 10^7$, Figure~\ref{fig:content3} illustrates that $4$ POD modes have the ability to attain $77.62\%$ of the RIC and $45$ modes are required to capture more than $99\%$ of the energy.   

%****************************************
\begin{figure}[htbp]
	\centering
	\includegraphics[scale=0.3]{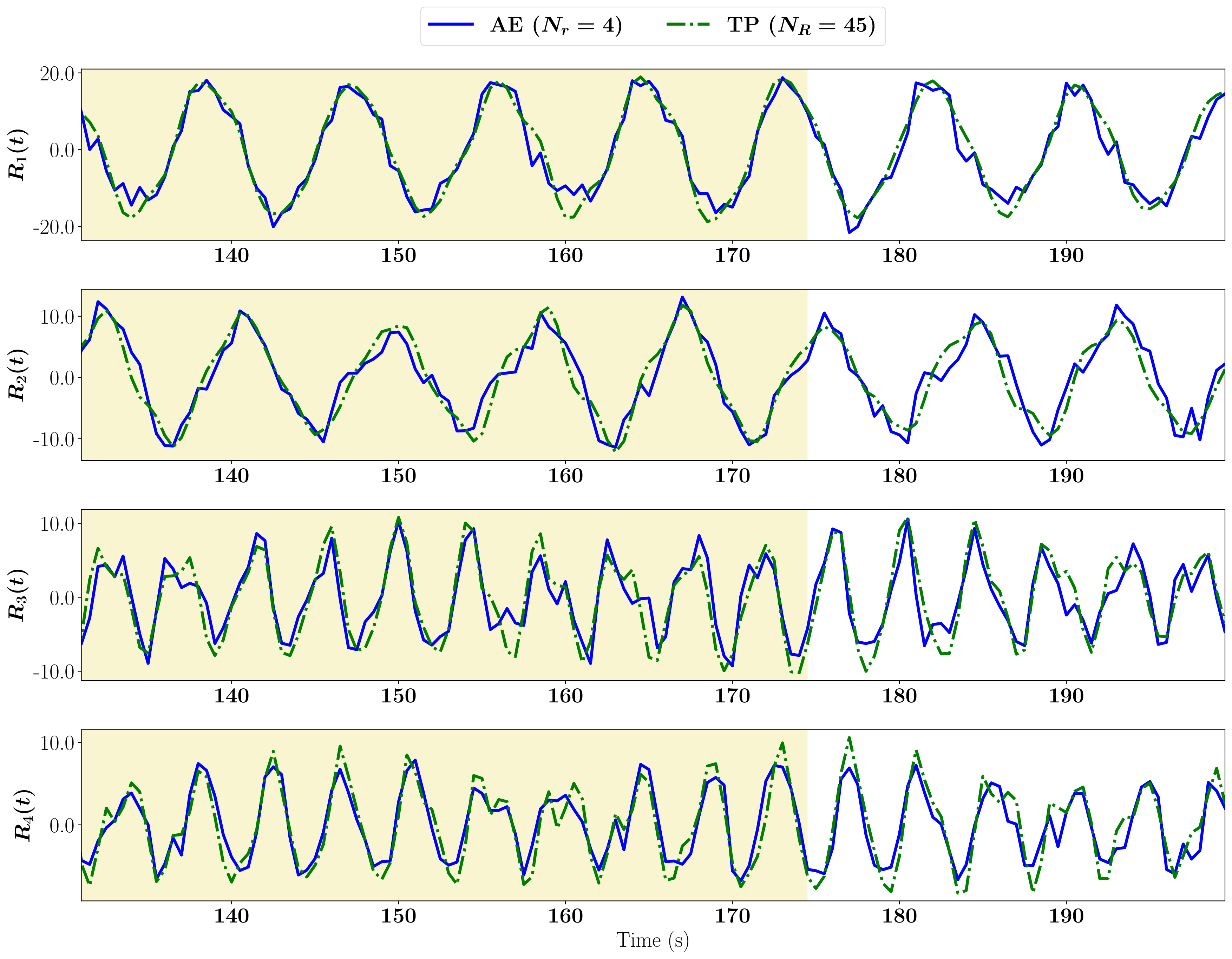}
	\caption{Variation of first four POD temporal coefﬁcients 
	%$\alpha_1,\;\alpha_2,\;\alpha_3,\;\alpha_4$
	for $\text{Ra}=1\times 10^{7}$.
	The training set for the AE network is from $t = 0$ to $t=175$ s.}
	\label{fig:fig12}
\end{figure}

%****************************************
Figure \ref{fig:fig12} presents the time variation of the first $4$ POD modes for $\text{Ra}=1\times 10^{7}$. Although the AE is able to reconstruct the $45$ POD modes with $4$ neurons in the latent space, there is a little discrepancy between AE and TP curves.
Besides, the LSTM network has to learn the curve that AE makes. Consequently, the more chaos curve AE makes the more arduous is for LSTM to follow the curve.
Having said that, to build NLPOD models for higher Ra numbers we need to focus on AE network too as the AE is going to be a bottleneck. Otherwise, we are not going to reduce data to a few modes.

%****************************************
\begin{figure}[htbp]
	\centering
	\includegraphics[scale=0.3]{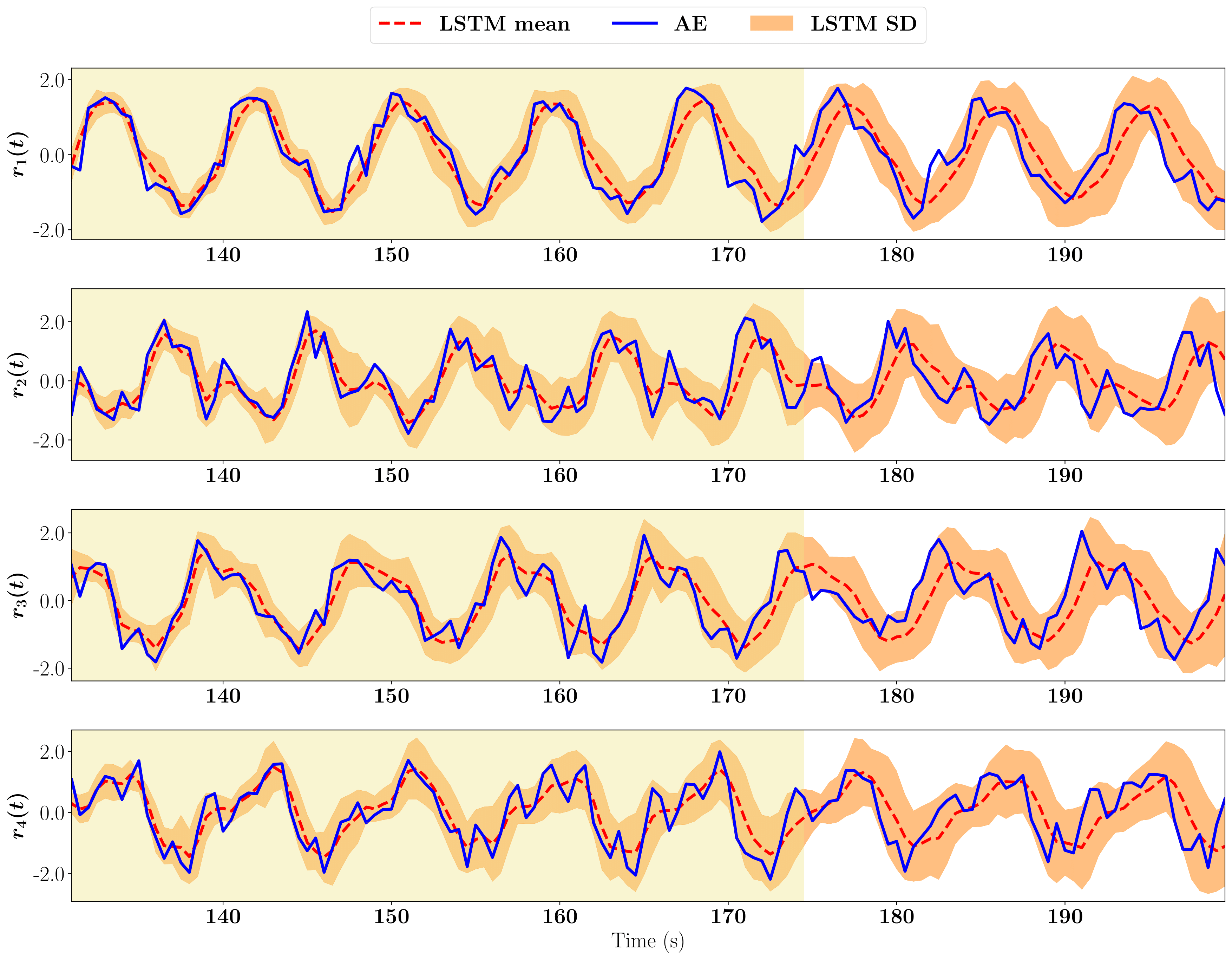}
	\caption{Evolution of autoencoder modes in latent space with the LSTM for $\text{Ra}=1\times 10^{7}$.
	The training set for the LSTM network is from $t = 0$ to $t=175$ s.}
	\label{fig:fig13}
\end{figure}

%****************************************
Figure \ref{fig:fig13} shows the time evolution of the first four AE modes for $\text{Ra}=1\times 10^7$.
The aim of this plot is to show how LSTM performs on predicting of the almost chaotic dynamics or irregular dynamic time series.
Although the LSTM can capture low frequency parts of the curve, its prediction is not able to exactly match the high frequency parts.
It also shows that in the extrapolatory region, there is a shift between the LSTM prediction and the AE prediction.
The $\text{SD}$ bars cover the AE curve. In other words, our 64 LSTM models can provide us with a bound that we are certain it covers the true or AE curve.

%****************************************
\begin{figure}[htbp]
	\centering
	\includegraphics[scale=0.3]{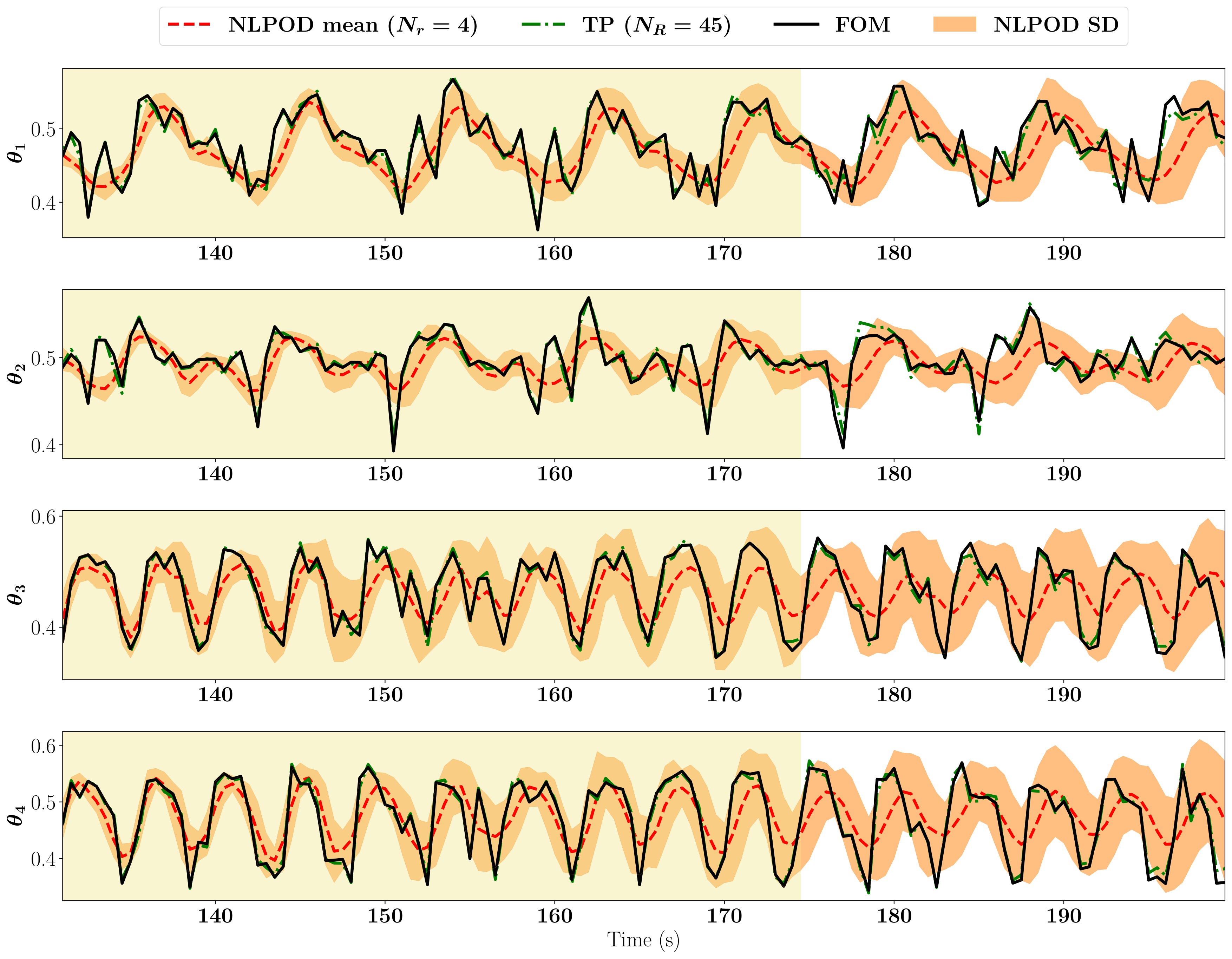}
	\caption{Temperature evolution in physical space for
	$\text{Ra}=1\times 10^{7}$ at four locations.}
	\label{fig:fig14}
\end{figure}

%****************************************
In Figure~\ref{fig:fig14}, TP is still very close to FOM curve since we remove uncorrelated data in two steps employing POD and AE instead of cutting all the modes from the mesh size to $4$ with only POD.
Figure~\ref{fig:fig14} where we observe the capability of the NLPOD to reconstruct trends for temperature field in the physical space from few modes.

%****************************************
%****************************************

\begin{figure}[htbp]
	\centering
	\includegraphics[scale=0.5]{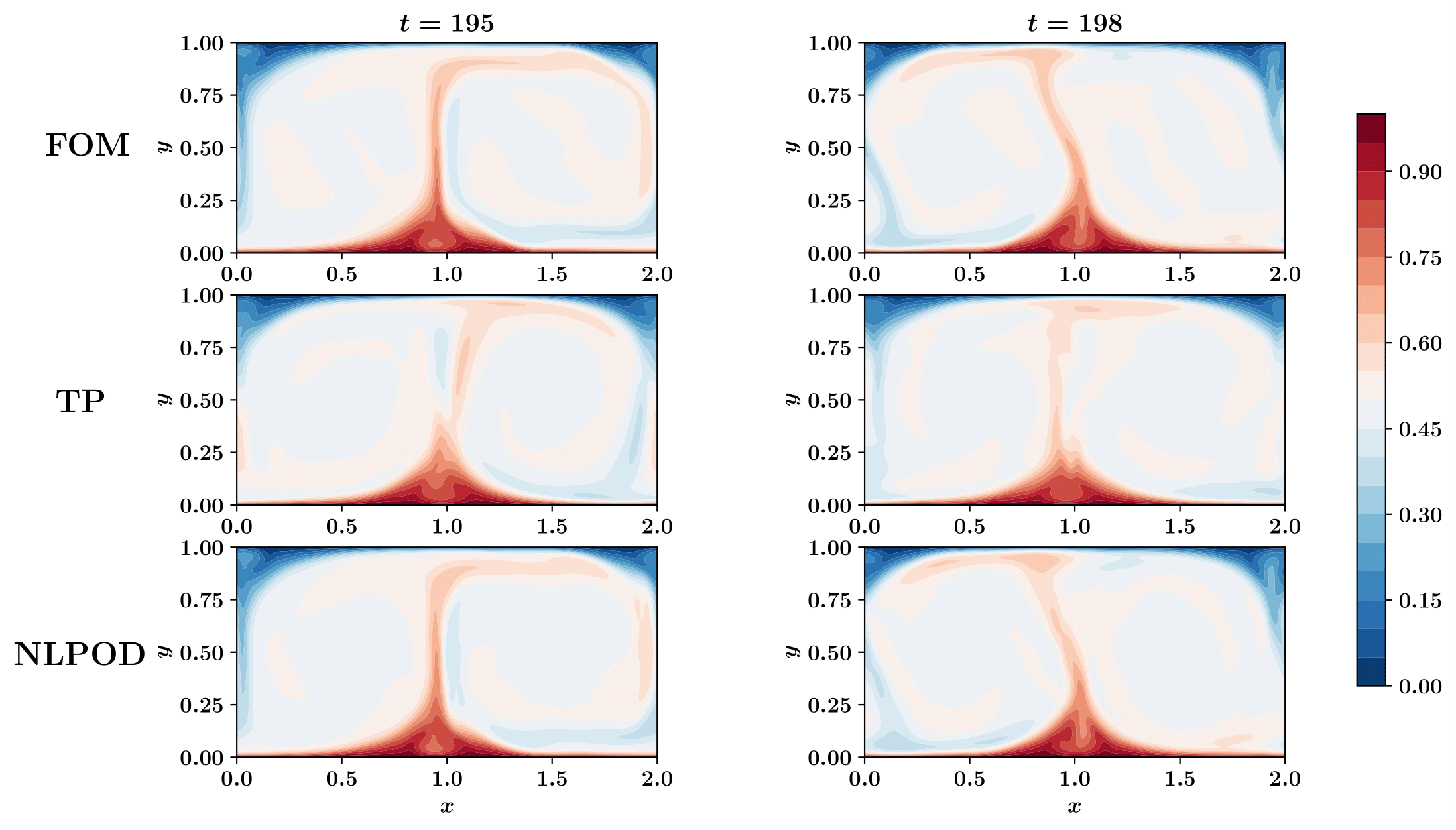}
	\caption{A comparison of temperature fields for $\text{Ra}=1\times 10^{7}$ at two different times.}
	\label{fig:fig15}
\end{figure}

Figure~\ref{fig:fig15} shows temperature contours at two different times.
The coherent structures of the temperature fields are reconstructed by the NLPOD compared with both TP and FOM.
%****************************************
%****************************************
%--------------------------------

%========================================================
\section{Concluding remarks}
\label{conclusions}
%========================================================
In this study, we investigate a nonintrusive reduced order model (ROM) for convection-dominated fluid flows
that can be used for other nonlinear continuum phenomena.
Our ROM, nonlinear proper orthogonal decomposition (NLPOD), is based on
a proper orthogonal decomposition to recognize dominant but linear patterns
from the data and to utilize an autoencoder technology with perceptron layers
for further reduction and nonlinear \textcolor{rev1}{pattern identification.}
Then, we use a long short term memory architecture to learn the dynamics of
low fidelity data in latent space.
This data-driven modeling approach relies on only stored data and can be applied without having access to the governing equations. 

\textcolor{rev1}{While the approach does not completely lift the Kolmogorov barrier, based on our analysis we found that the NLPOD significantly reduces the number of degrees of freedom needed to represent the underlying dynamics. Indeed, the degree of irregularity increases as the Rayleigh number rises, producing a more chaotic or turbulent spatiotemporal dynamics. We also highlight that the full order model dimension can be on the order of billions (and even higher) for three-dimensional turbulent industrial or geophysical flows. In those cases, the NLPOD approach might still reduce the ROM dimension to a handful of latent variables if there are spatiotemporal structures. However, while the Kolmogorov n-width increases, the models' uncertainty grows as shown in our results. By defining feasible maps between the observation space and the latent variables, a viable method for minimizing the uncertainty of such ROM solutions might be to use nonlinear filtering or dynamic data assimilation techniques, a topic we plan to address in a subsequent paper.}

%However, we use a set of partial differential equations and a fourth order numerical scheme to generate data.
%Add one more paragraph
%In the future, we plan on applying our method to unstructured grids.
%We also plan to employ it to calculate a cost function for the purpose of shape optimization.

% \section*{Acknowledgements}

% An unnumbered section, e.g.\ \verb"\section*{Acknowledgements}", may be used for thanks, etc.\ if required and included \emph{in the non-anonymous version} before any Notes or References.

\section*{Data availability}
The data that support the findings of this study
are available within the article. The datasets used and/or analysed during the current study are available from the corresponding author on reasonable request. The implementation and open-source Python codes are available at \url{https://github.com/Saeed-Akbari/POD-AE-LSTM}.

\section*{Disclosure statement}

We declare we have no competing interests.

\section*{Funding}

This material is based upon work supported by the U.S. Department of Energy, Office of Science, Office of Advanced Scientific Computing Research under Award Number DE-SC0019290. O.S. gratefully acknowledges their support.%O.S. gratefully acknowledges their Early Career Research Program support.

Disclaimer: This report was prepared as an account of work sponsored by an agency of the United States Government. Neither the United States Government nor any agency thereof, nor any of their employees, makes any warranty, express or implied, or assumes any legal liability or responsibility for the accuracy, completeness, or usefulness of any information, apparatus, product, or process disclosed, or represents that its use would not infringe privately owned rights. Reference herein to any specific commercial product, process, or service by trade name, trademark, manufacturer, or otherwise does not necessarily constitute or imply its endorsement, recommendation, or favoring by the United States Government or any agency thereof. The views and opinions of authors expressed herein do not necessarily state or reflect those of the United States Government or any agency thereof.

% \section*{Notes on contributors}
% %A.B. and B.C. carried out the experiment. A.B. wrote the manuscript ... All authors discussed the results and contributed to the final manuscript.
% O.S. conceived the research and supervised the study. S.A. performed the numerical experiments, analysed the data, and wrote the initial draft of the manuscript.  O.S. and S.P. provided feedback and contributed in shaping the research, analysis and manuscript. All coauthors reviewed the manuscript, discussed the results, and contributed to the final manuscript.

%All authors contributed equally to this work.  

% \section*{Nomenclature/Notation}

% An unnumbered section, e.g.\ \verb"\section*{Nomenclature}" (or \verb"\section*{Notation}"), may be included if required, before any Notes or References.

% \section*{Notes}

% An unnumbered `Notes' section may be included before the References (if using the \verb"endnotes" package, use the command \verb"\theendnotes" where the notes are to appear, instead of creating a \verb"\section*").

\bibliographystyle{apacite}
\bibliography{references}

\end{document}